\RequirePackage{fix-cm}
\documentclass[twocolumn,epjc3]{svjour3} 
\usepackage[separate-uncertainty=true]{siunitx}
\DeclareSIUnit{\countspersecond}{cps}
\DeclareSIUnit\bar{bar}
\usepackage{graphicx}
\usepackage{mathtools}
\PassOptionsToPackage{
  backend=bibtex8,
  bibencoding=ascii,
  language=auto,
  style=numeric,
  giveninits=true,
  citestyle=numeric-comp,
  isbn=false,
  url=false,
  sorting=none,
  maxbibnames=2,
  natbib=true}{biblatex}
\usepackage{biblatex}
\AtEveryBibitem{\clearfield{note}}
\AtEveryBibitem{\clearfield{month}}
\AtEveryBibitem{\clearfield{issue}}
\renewbibmacro{in:}{} 
\DefineBibliographyStrings{english}{
    andothers = {\em et\addabbrvspace al\adddot}
}
\renewbibmacro*{doi+eprint+url}{%
    \printfield{doi}%
    \newunit\newblock%
    \iftoggle{bbx:eprint}{%
        \usebibmacro{eprint}%
    }{}%
    \newunit\newblock%
    \iffieldundef{doi}{%
        \usebibmacro{url+urldate}}%
        {}%
    }

\DeclareFieldFormat{pages}{} 

\smartqed  
\RequirePackage{graphicx}
\RequirePackage[colorlinks,citecolor=blue,urlcolor=blue,linkcolor=blue]{hyperref}
\RequirePackage{textgreek}
\RequirePackage{amsmath}
\RequirePackage{amssymb}
\RequirePackage{multirow}
\RequirePackage{floatrow}
\RequirePackage{subfig}
\RequirePackage{booktabs}
\RequirePackage{placeins}
\RequirePackage{varwidth}
\RequirePackage{braket}
\RequirePackage{hyperref}

\usepackage{wasysym} 
\usepackage{xcolor}
\definecolor{C0}{HTML}{1F77B4}
\definecolor{C1}{HTML}{ff7f0e}
\definecolor{C2}{HTML}{2ca02c}
\definecolor{C3}{HTML}{d62728}

\DeclarePairedDelimiterXPP\BigOSI[2]
  {\mathcal{O}}{(}{)}{}%
  {\SI{#1}{#2}}

\journalname{}

\begin{document}
\sloppy

\title{Development of a Silicon Drift Detector Array to Search for keV-scale Sterile Neutrinos with the KATRIN Experiment}

\author{D.~Siegmann\thanksref{addr1,addr2,e1} \and
F.~Edzards\thanksref{addr1,addr2} \and
C.~Bruch\thanksref{addr1,addr2} \and
M.~Biassoni\thanksref{addr3,addr4} \and
M.~Carminati\thanksref{addr3,addr5} \and
M.~Descher\thanksref{addr6} \and
C.~Fiorini\thanksref{addr3,addr5} \and
C.~Forstner\thanksref{addr1,addr2} \and
A.~Gavin\thanksref{addr7} \and
M.~Gugiatti\thanksref{addr3,addr5} \and
R.~Hiller\thanksref{addr6}\and
D.~Hinz\thanksref{addr6} \and
T.~Houdy\thanksref{addr8} \and
A.~Huber\thanksref{addr6} \and
P.~King\thanksref{addr3,addr5} \and
P.~Lechner\thanksref{addr9} \and
S.~Lichter\thanksref{addr6} \and
D.~Mießner\thanksref{addr9} \and
A.~Nava\thanksref{addr3,addr4} \and
A.~Onillon\thanksref{addr1} \and
D.~C.~Radford\thanksref{addr10} \and
D.~Spreng\thanksref{addr1,addr2} \and
M.~Steidl\thanksref{addr6} \and
P.~Trigilio\thanksref{addr11} \and
K.~Urban\thanksref{addr1,addr2} \and
D.~Vénos\thanksref{addr12} \and
J.~Wolf\thanksref{addr6} \and
S.~Mertens\thanksref{addr1,addr2}
}

\thankstext{e1}{E-mail: daniel.siegmann@tum.de}
\institute{Technical University of Munich, TUM School of Natural Sciences, Physics Department, 85748 Garching, Germany \label{addr1} \and
Max Planck Institute for Physics, Boltzmannstr. 8, 85748 Garching, Germany \label{addr2} \and
INFN, Sezione di Milano, Via Celoria 16, 20133 Milano, Italy \label{addr3} \and
Universita degli Studi di Milano-Bicocca, Piazza dell'Ateneo Nuovo, 1-20126, Milano, Italy \label{addr4} \and
DEIB, Politecnico di Milano, Via Golgi 40, 20133 Milano, Italy \label{addr5} \and
IAP, Karlsruhe Institute of Technology, Hermann-von-Helmholtz-Platz 1, 76344 Eggenstein-Leopoldshafen, Germany \label{addr6} \and
University of North Carolina at Chapel Hill, 120 E.~Cameron Avenue, Chapel Hill, 27599 North Carolina, USA \label{addr7} \and
IJCLab, Université Paris-Saclay, 15 rue Georges Clémenceau, 91405 Orsay cedex, France \label{addr8} \and
Halbleiterlabor der Max-Planck-Gesellschaft, Isarauenweg, 85748 Garching, Germany \label{addr9} \and 
Oak Ridge National Laboratory, 1 Bethel Valley Rd, Oak Ridge, TN 37830, USA \label{addr10} \and
XGLab SRL, Bruker Nano Analytics, Via Conte Rosso 23, 20134 Milano, Italy \label{addr11} \and 
Nuclear Physics Institute of the CAS, v.~v.~i., 250 68 \v{R}e\v{z}, Czech Republic \label{addr12} 
}

\date{Received: date / Accepted: date}

\maketitle

\setcounter{tocdepth}{4}

\begin{abstract}
Sterile neutrinos in the keV mass range present a viable candidate for dark matter. They can be detected through single~\textbeta~decay, where they cause small spectral distortions. The Karlsruhe Tritium Neutrino~(KATRIN) experiment aims to search for keV-scale sterile neutrinos with high sensitivity. To achieve this, the KATRIN beamline will be equipped with a novel multi-pixel silicon drift detector focal plane array named TRISTAN. In this study, we present the performance of a TRISTAN detector module, a component of the eventual 9-module system. Our investigation encompasses spectroscopic aspects such as noise performance, energy resolution, linearity, and stability.
\end{abstract}

\section{Introduction}\label{ch:introduction}
\begin{figure}[!t]
\centering
\includegraphics[width=1.0\linewidth]{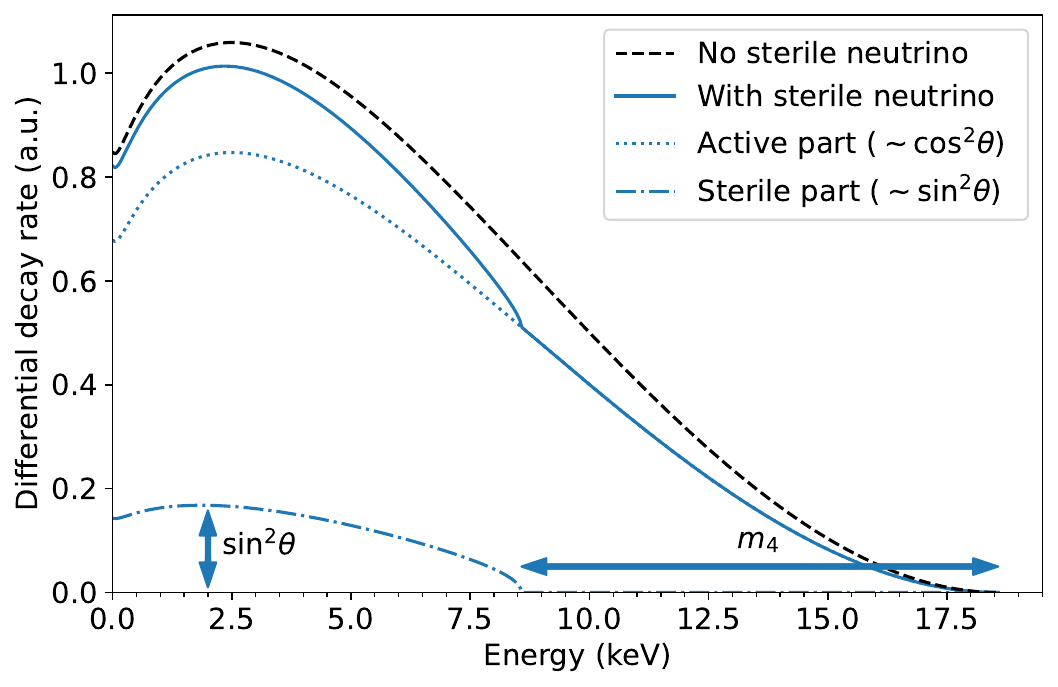}
\caption{Electron energy spectrum of tritium \textbeta-decay shown with and without the imprint of a sterile neutrino. For illustrative purposes, a neutrino mass of~\mbox{$m_4=\SI{10}{\kilo\electronvolt}$} and an unphysically large mixing amplitude of~$\sin^2\theta=0.2$ have been used.}
\label{fig:steril_neutrino}
\end{figure}

Sterile neutrinos\footnote{In the following, the term~\textit{sterile neutrino} denotes an additional fourth mass eigenstate which is not purely sterile and can thus mix with the three active neutrinos.} are a natural and viable extension to the Standard Model of particle physics~\cite{Adhikari_2017}. If their mass is in the keV~range, they are a suitable dark matter candidate~\cite{Adhikari_2017,PhysRevLett.72.17,PhysRevLett.82.2832,BOYARSKY20191}. Depending on their production mechanism, they can act effectively as cold or warm dark matter. This can potentially help to mitigate tensions between predictions of pure cold dark matter scenarios and the observations of small scale structures in our Universe~\cite{Murgia_2017}. Indirect searches and cosmological observations set stringent limits on the active-sterile neutrino mixing amplitude of~\mbox{$10^{-6}>\sin^2\theta>10^{-10}$} in a mass range of~$\SI{1}{\kilo \electronvolt}<m_4<\SI{50}{\kilo \electronvolt}$~\cite{Boyarsky_2009,Narayanan_2000,PhysRevLett.97.191303,PhysRevD.16.1444,PhysRevD.25.766}. However, these limits are model-dependent and can be relaxed by several orders of magnitude by modifying the dark matter decay models~\cite{PhysRevD.100.115035}. Current laboratory limits on the mixing amplitude with values of~\mbox{$\sin^2\theta>10^{-4}$} are orders of magnitude weaker compared to the cosmological limits~\cite{Aker_2022,Abdurashitov2017,PhysRevC.36.1504,PhysRevC.32.2215,OHI1985322,PhysRevD.47.4840,Mueller_1994,HOLZSCHUH1999247}. 
\\\\
One method to search for sterile neutrinos in a laboratory-based experiment is via their production in single \textbeta-decay. The electron-flavor neutrino emitted in the decay is a superposition of the neutrino mass eigenstates, including the keV-scale sterile neutrino. Correspondingly, the measured electron energy spectrum, shown in figure~\ref{fig:steril_neutrino}, is a superposition of the active and sterile neutrino decay branches. The experimental signature is a spectral distortion at energies~$E\leq E_0-m_4$, which includes a kink-like feature. Here,~$E_0$ denotes the kinematic endpoint of the decay. A search based on single \textbeta-decay is sensitive to sterile neutrino masses of~$m_4\leq E_0$.
%
\\\\
The \mbox{Karlsruhe} Tritium Neutrino~(KATRIN) experiment is designed to measure the effective electron antineutrino mass with a sensitivity better than~\SI{300}{\milli\electronvolt}, by performing a measurement of the tritium \textbeta-decay spectrum near its endpoint energy of $E_0=\SI{18.6}{\kilo\electronvolt}$~\cite{Angrik_2005,Aker_2021}. Recently, the first sub-eV limit on $m_{\nu}$ of~\SI{0.8}{\electronvolt}~\mbox{(90\% CL)} based on the first two high-activity tritium measurement campaigns has been published~\cite{PhysRevLett.123.221802, PhysRevD.104.012005, Katrin_Nature}. The neutrino mass program of KATRIN is foreseen to continue until the end of~2025. Subsequently, the collaboration plans to upgrade the beamline, in order to perform a keV-scale sterile neutrino search by investigating the entire tritium \textbeta-decay spectrum. Considering the high source activity of KATRIN~(\SI{e11}{\becquerel}), it would be in principle possible to achieve a statistical sensitivity to keV-scale sterile neutrinos in the order of~$\sin^2\theta<10^{-8}$ within~\SI{3}{years} of data taking~\cite{Mertens_2015}. However, due to detector rate limitations and systematic uncertainties, the collaboration explores the possibility of reaching a sensitivity at the level of~$10^{-6}$.
\\\\
As opposed to the neutrino mass measurements where the region of interest extends \SI{40}{\electronvolt} below the endpoint energy $E_0$, the signature of a keV-scale sterile neutrino lies much further away from the endpoint, due to its unconstrained mass scale. By extending the measurement interval deeper into the tritium spectrum, the KATRIN experiment has the potential to search for sterile neutrino masses of~$m_4\leq E_0 \approx \SI{18.6}{\kilo\electronvolt}$. The current focal plane detector system~\cite{Angrik_2005,Aker_2021} is not designed to handle the associated exceedingly high count rates. To cope with this challenge, a new detector system, named TRISTAN, with more than~\num{1000}~pixels and a targeted energy resolution of \SI{300}{\electronvolt} at \SI{20}{\kilo\electronvolt} is currently being developed. It is based on the silicon drift detector~(SDD) technology which provides excellent spectroscopic properties at the envisioned high rates of~\SI{100}{\kilo \countspersecond} per pixel. Thanks to their small readout anode, and thus small detector capacitance in the order of \SI{120}{\femto\farad}, SDDs feature an excellent energy resolution of~\SI{180}{\electronvolt}~FWHM for \SI{5.9}{\kilo\electronvolt} x-rays at short energy filter shaping times of \SI{0.5}{\micro\second}~\cite{2004_Lechner}. 
\begin{figure}[!t]
\subfloat[Baseline (9 modules)\label{graph:tristan_phase1}]{%
\includegraphics[width=0.48\textwidth]{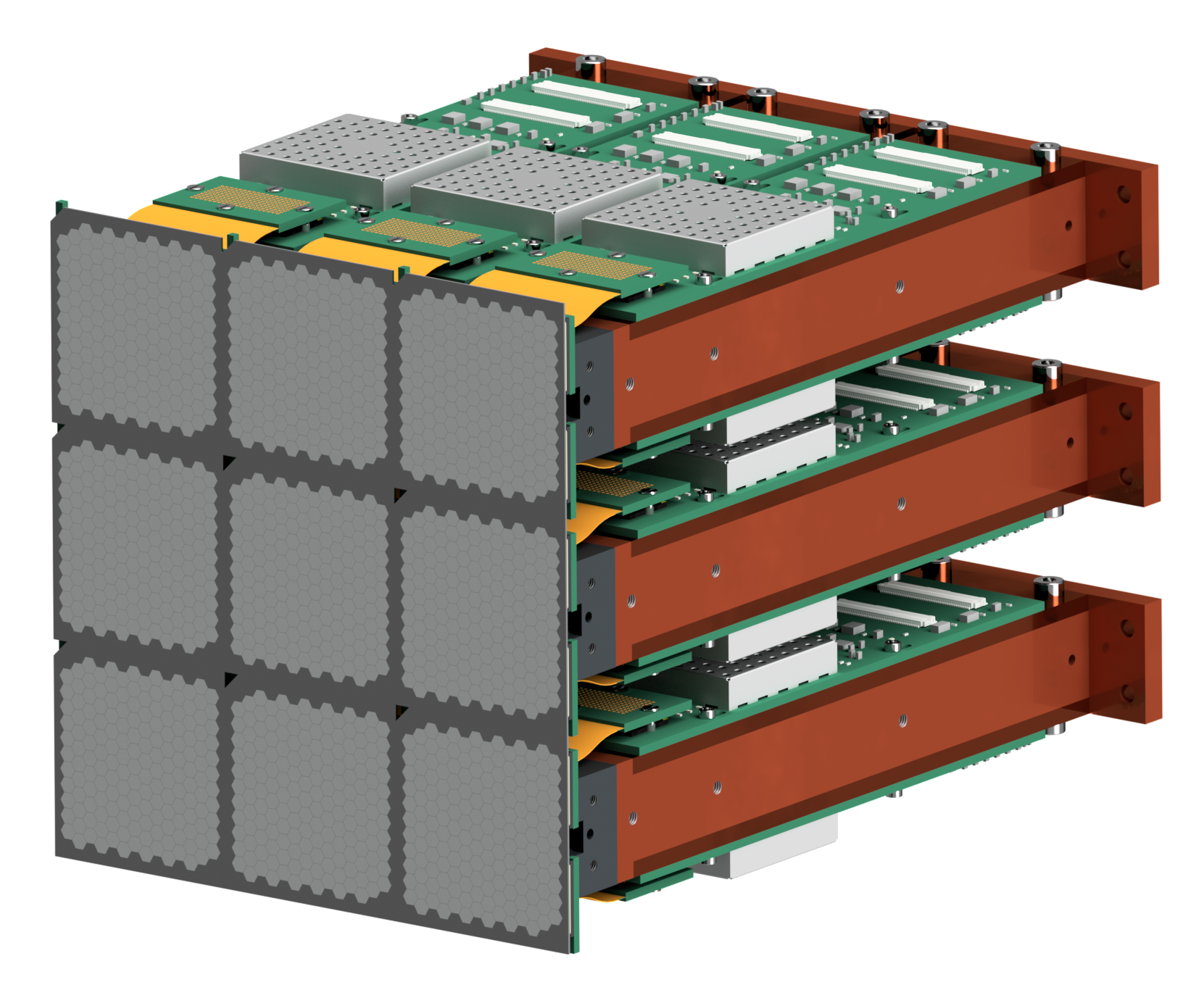}}
\quad
\subfloat[Upgrade (21 modules)\label{graph:tristan_phase2}]{%
\includegraphics[width=0.48\textwidth]{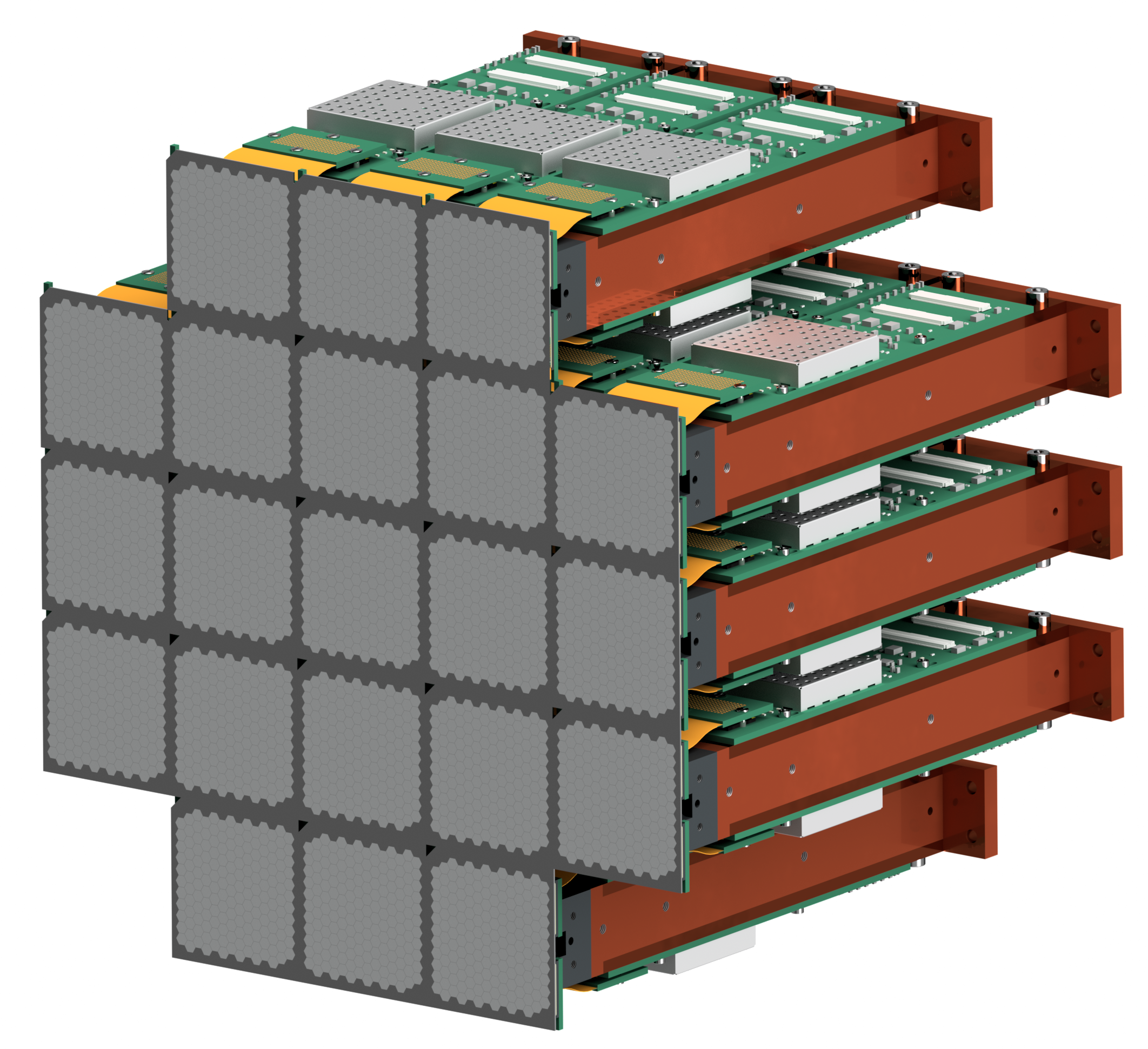}}
\caption{Renderings of the TRISTAN detector system consisting of 9~modules (baseline design) and a potential upgrade consisting of 21~modules.}
\label{fig:TRISTAN_detector}
\end{figure}
%
\\\\
The TRISTAN detector system features a modular design. It consists of several identical detector modules that can be mounted next to each other as shown in figure~\ref{fig:TRISTAN_detector}. Each detector module has~\SI{166}{pixels}. In the baseline design, the detector system will be composed  of~\num{9}~detector modules with a total of about~\num{1500}~pixels. In this configuration, the detector system can be installed in the beamline of the KATRIN experiment with only minor modifications of the current detector section. Due to the modular design of the detector system, a potential upgrade using~21 modules, is technologically feasible and could further improve the statistical sensitivity. However, due to the bigger dimensions, such an upgrade would require a redesign of the detector section.
\\\\
The main goals of this work are the commissioning and the characterization of the first TRISTAN detector modules with photons and electrons. We present results obtained in dedicated test stands and the first operation of a detector module under quasi-realistic conditions~(e.g.~vacuum, electric and magnetic fields) in the KATRIN monitor spectrometer. 

\section{TRISTAN detector module}\label{ch:tristan_det}
\begin{figure}[!t]
\centering
\includegraphics[width=0.9\linewidth]{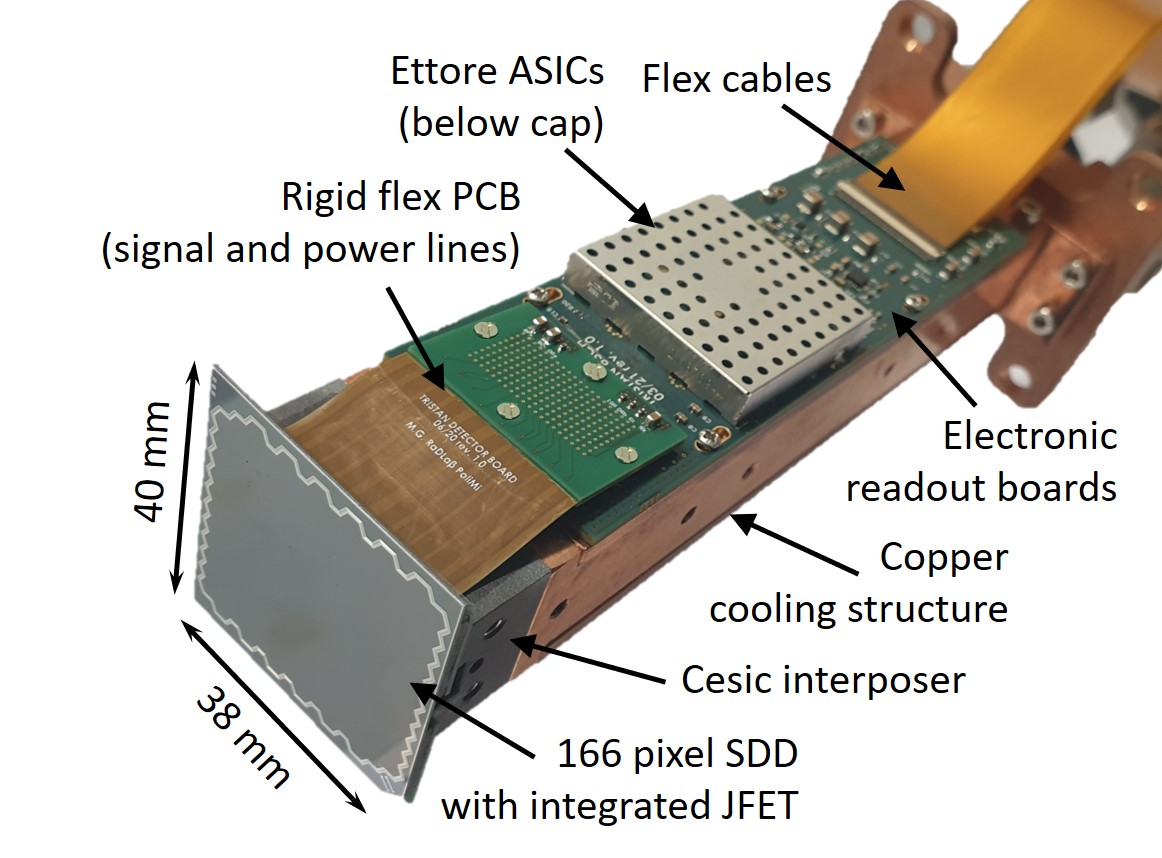}
\caption{Photograph of a TRISTAN detector module. It consists of a monolithic \num{166}~pixel SDD chip which is glued on a Cesic interposer. The interposer is connected to a copper cooling structure. The readout electronics boards with dedicated readout chips are mounted on the top and bottom surfaces of the copper block.}
\label{fig:module}
\end{figure}

The TRISTAN detector module shown in figure~\ref{fig:module} hosts of a large rectangular monolithic multi-pixel SDD chip which is glued on a silicon ceramic composite called Cesic and is connected to a copper cooling structure. The requirements for the detector module and its individual components are explained in the following sections.

\subsection{Detector module requirements}
To cover the entire electron flux tube of the KATRIN experiment, a detector system with a diameter of approximately~\SI{20}{\cm} is required. Even though the concept of the SDD is very flexible in terms of size and shape, building one large monolithic SDD chip with more than~\SI{1000}{pixels} is considered not to be technologically feasible. Therefore, the detector system has to be segmented into smaller detector modules. As shown in figure~\ref{fig:module}, the entire on-module electronics are placed in a plane perpendicular to the SDD chip, on two identical electronics readout boards. This arrangement allows for an almost seamless placement of the detector modules next to each other, see~figure~\ref{fig:TRISTAN_detector}, minimizing the insensitive area of the detector system.
\\\\
All components of the detector module have been selected carefully to ensure compatibility with the environment in the KATRIN experiment, that is based on the MAC-E filter~(magnetic adiabatic collimation with an electrostatic filter) principle~\cite{PICARD1992345,Angrik_2005,Aker_2021}. In particular, they need to be suited for high magnetic fields of up to~\SI{2.5}{\tesla}~\cite{Angrik_2005,Aker_2021}, and need to have a low outgassing rate since the detector chamber is connected to the MAC-E filter vessel, called main spectrometer. For the ongoing neutrino mass measurements, an ultra high vacuum of~\SI{e-11}{\milli\bar} in the main spectrometer is required to keep the background rate at a level of~\SI{220}{\milli\countspersecond}~\cite{Katrin_Nature}. To achieve this pressure, the gas load stemming from the detector section has to be on the order of~\mbox{\SI{5e-8}{\milli\bar\liter\per\second}}~\cite{2022_Tristan_Design_Report}. For the sterile neutrino search with KATRIN, the requirements on the vacuum are less stringent. Compared to the neutrino mass measurement mode close to the endpoint, the signal rate in the full spectrum is more than eight orders of magnitude higher than the current background rate. Therefore, a higher background level can be tolerated which relaxes the vacuum level requirements. The dominant outgassing contribution of the TRISTAN detector system can be attributed to the material of the electronics PCBs~(FR4) with a value of about~\SI{3e-7}{\milli\bar\liter\per\second} for each detector module~\cite{2022_Tristan_Design_Report}. Since the width of the electronics PCBs is limited by the SDD chip size, a high-density 12-layer design is necessary to accommodate all required connections. Due to the complexity of the electronics PCBs, no other low-outgassing material could be used. This includes for example the polyimide Kapton which is used for the other electronic components. To compensate for the additional gas load stemming from the electronics PCBs, the conductance between the detector section and the main spectrometer has to be reduced. This is realized via a stainless steel grid which covers the gaps between the SDD chips. In addition, the pumping speed inside the detector section will be increased.

\subsection{SDD chip and readout electronics}
The monolithic SDD chip consists of 166~hexa\-gonal pixels. It is manufactured by the Semiconductor Laboratory~(HLL) of the Max Planck Society from a silicon wafer with a thickness of about~\SI{450}{\micro\meter}. SDDs are semiconductor detectors that use the principle of sideward depletion~\cite{Lechner_1996,Lechner_2001}. The detector bulk is depleted by applying a bias voltage to the entrance window side, i.e.~the side on which the radiation enters the detector. Several drift rings on the opposite side, the readout side, form the electric field within the detector. As soon as radiation enters the detector, electron hole pairs are created. The electrons are collected at a small anode~(diameter of~\SI{82}{\micro\meter}) on the readout side. In order to minimize detector-related effects such as charge sharing, backscattering, and back-reflection, a pixel size with a circumscribed diameter of~\SI{3.298}{\milli \metre} has been chosen~\cite{PhDKorzeczek_2020}. The area of the hexagon corresponds to a circle with a diameter of~\SI{3}{\milli \meter}. The pixels are arranged in a gapless and continuous way, resulting in a sensitive area of about~\SIrange[range-phrase={\,x\,},range-units=single]{37}{37}{\milli\meter\squared} and a total chip size of~\SIrange[range-phrase={\,x\,},range-units=single]{38}{40}{\milli\meter\squared}. 
\\\\
To take full advantage of the small anode  with a detector capacitance of only~\SI{180}{\femto\farad}, an n-channel junction-gate field-effect transistor~(nJFET) is directly integrated into each pixel. This configuration reduces the stray capacitance between the sensor and the charge sensitive amplifier~(CSA) and makes it possible to place the readout electronics several centimeters away from the detector. The signals are amplified by a readout application-specific integrated circuit~(ASIC) called ETTORE, that was specifically developed for the TRISTAN project~\cite{Trigilio_2018}. Each ETTORE ASIC provides the amplification for up to 12 pixels. The connection between the detector and the readout electronics is realized via wire bonds that are connected to a partially flexible PCB~(rigid flex PCB) that is mounted on the electronics board. The rigid flex cables allow for a~\SI{90}{\degree} bent connection between the detector plane and the plane of the electronics boards. More details on the readout electronics of the TRISTAN detector module can be found in references~\cite{Gugiatti_2021, PhDGugiatti_2022,Carminati_2023}.

\subsection{Assembly procedure}
In the following paragraphs, the different steps of the mounting procedure of the TRISTAN detector module will be described.
\begin{figure*}[!t]
\subfloat[Glue on Cesic\label{graph:assembly_cesic}]{\includegraphics[width=0.22\textwidth]{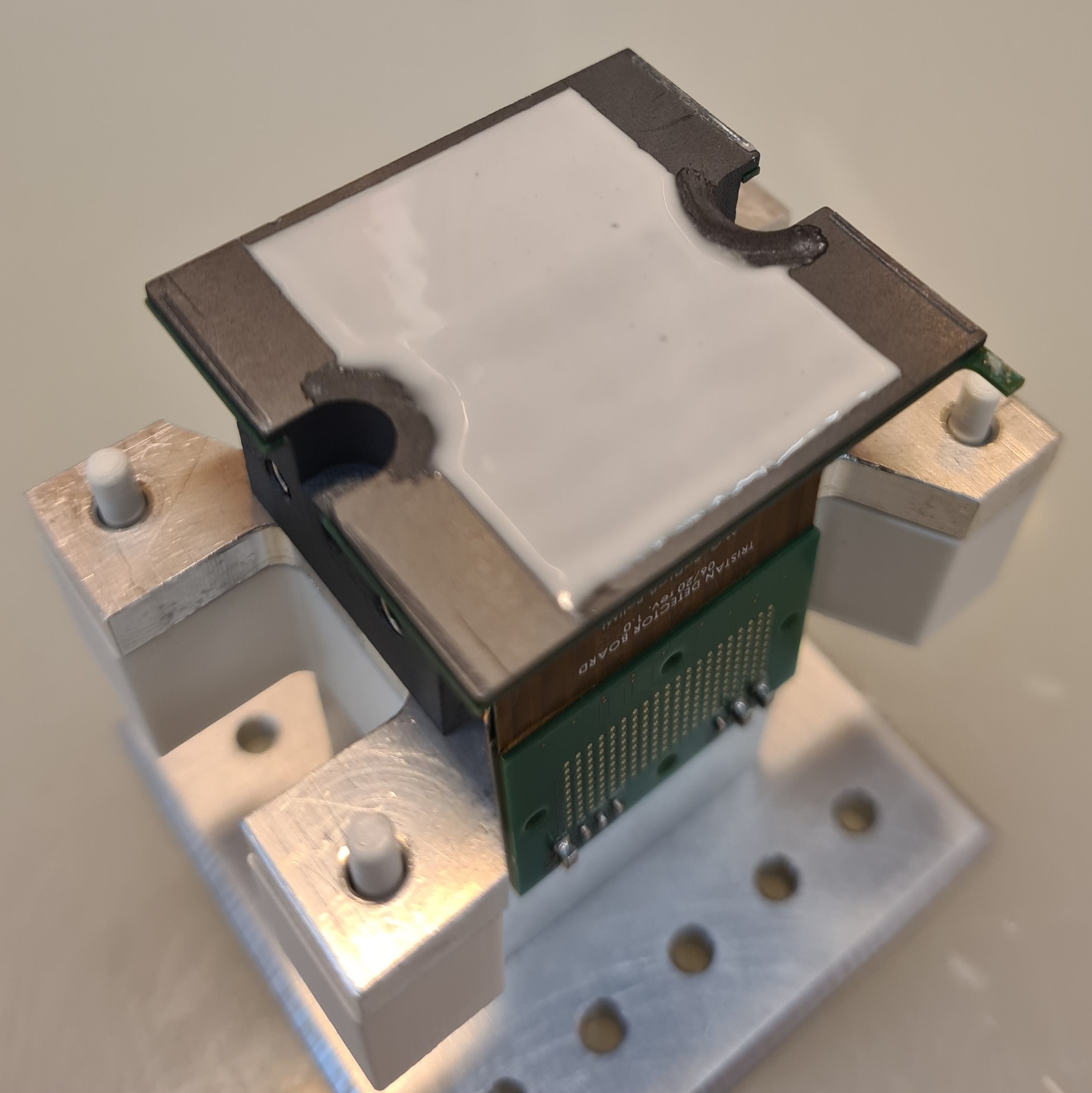}}\quad
\subfloat[SDD in PEEK jig\label{graph:assembly_sdd}]{\includegraphics[width=0.22\textwidth]{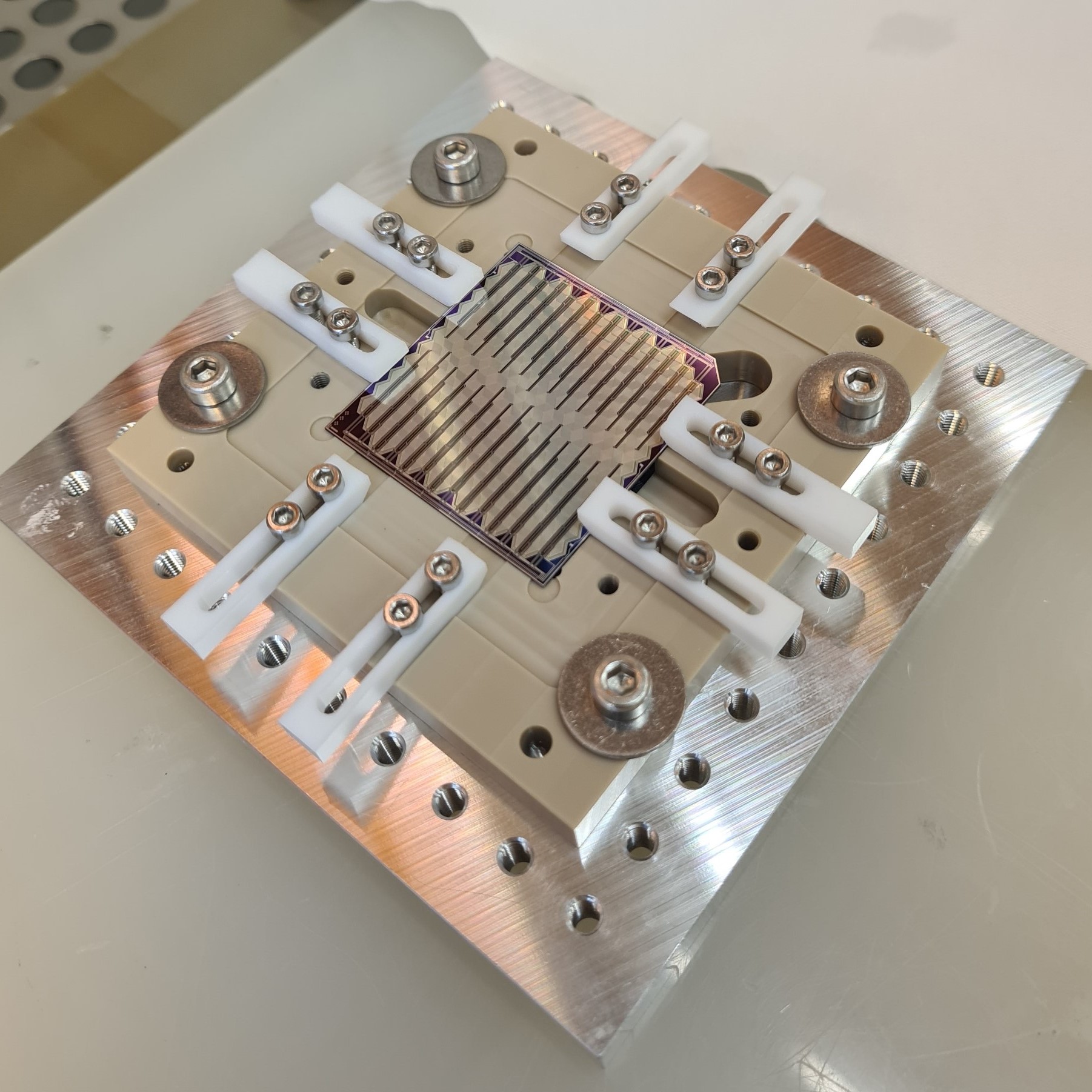}}\quad
\subfloat[Gluing\label{graph:assembly_all}]{\includegraphics[width=0.22\textwidth]{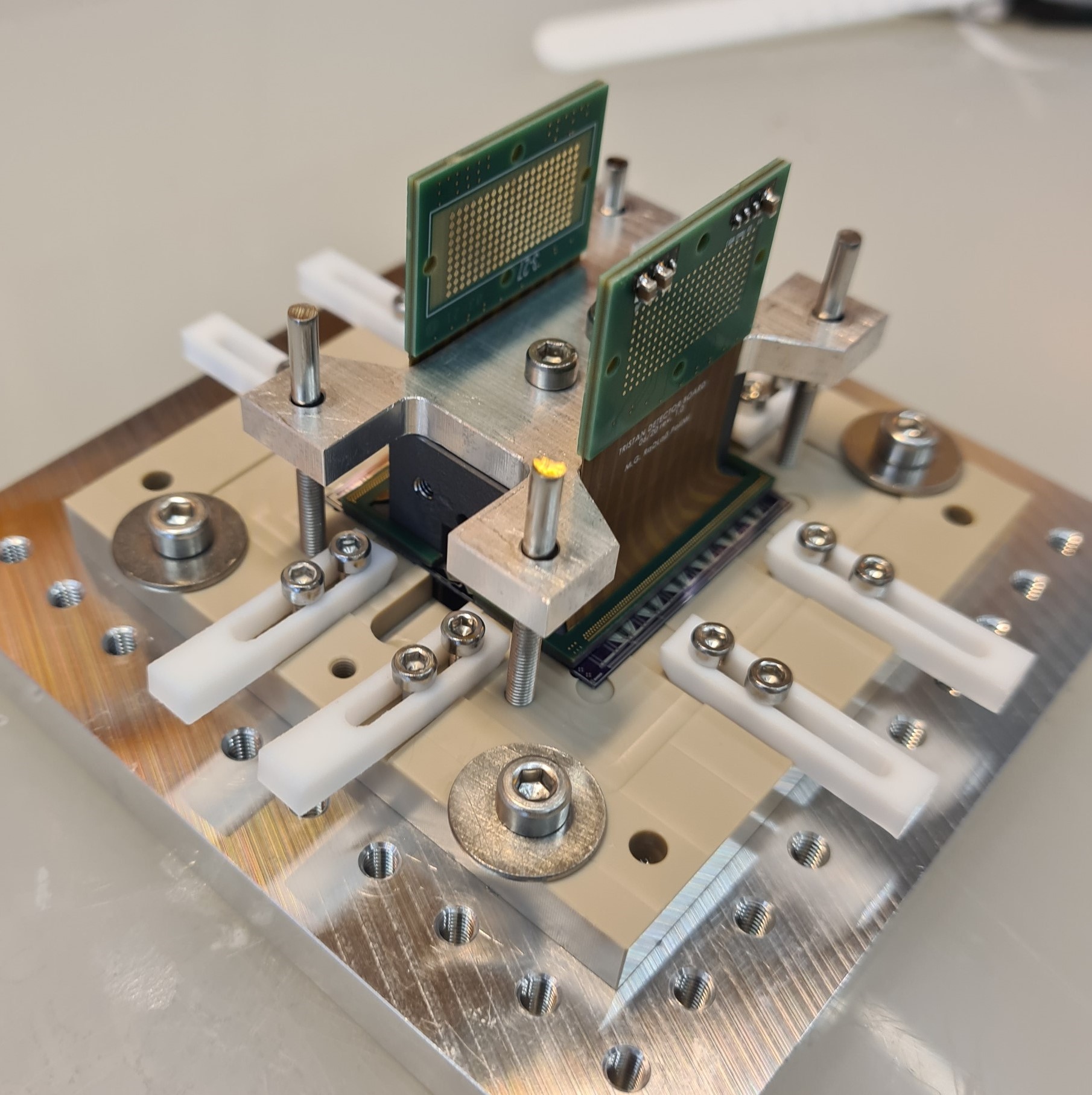}}\quad
\subfloat[Wire bonding\label{graph:assembly_bonding}]{\includegraphics[width=0.22\textwidth]{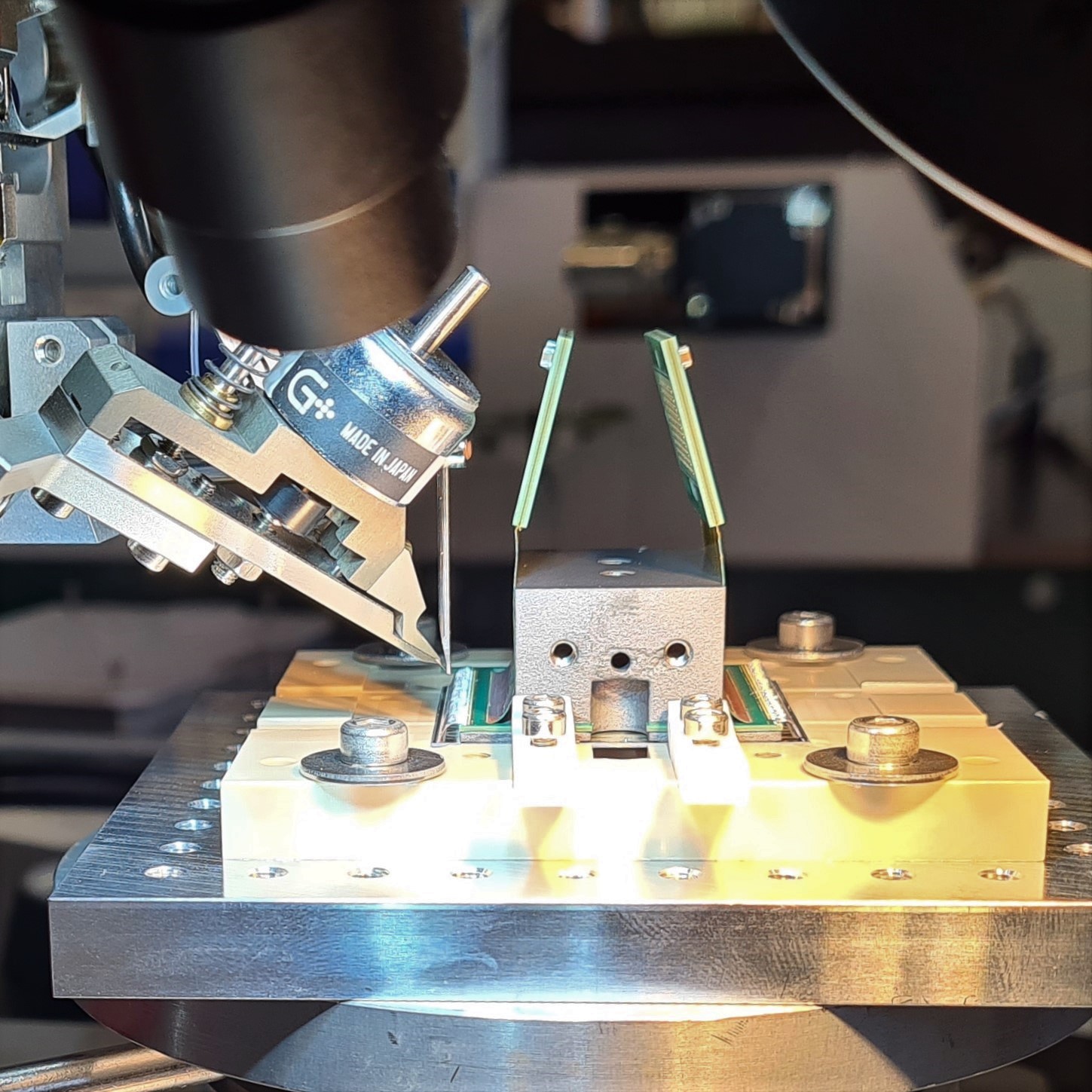}}\quad
\subfloat[Completed detector\label{graph:assembly_final}]{\includegraphics[width=0.22\textwidth]{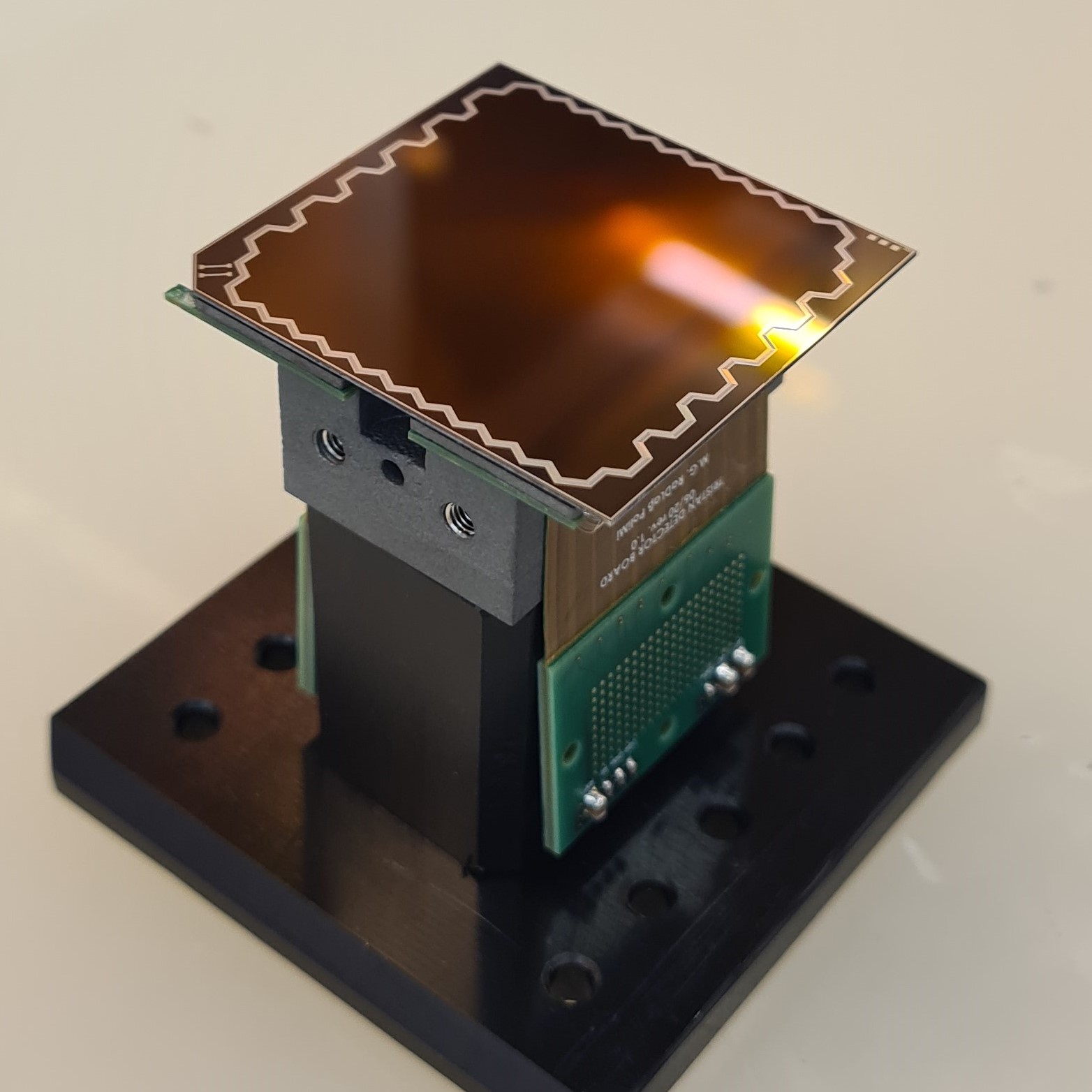}}\quad
\subfloat[Cross-section of the assembly\label{graph:assembly_schematic}]{\includegraphics[width=0.48\textwidth]{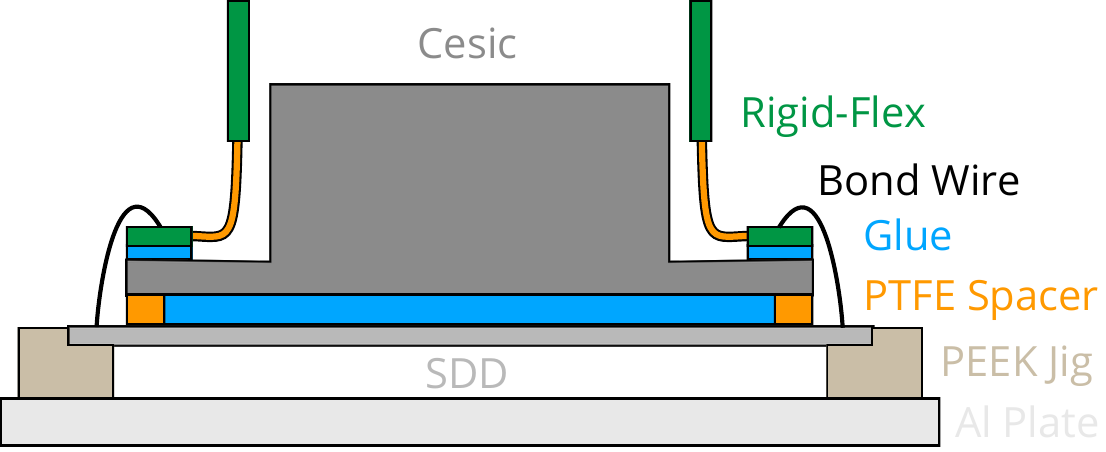}}
\caption{Photographs and schematic of the detector module assembly procedure. \protect\subref{graph:assembly_cesic}~Cesic interposer with white glue just before the assembly on the SDD chip. \protect\subref{graph:assembly_sdd}~SDD chip with the readout side facing up in the PEEK jig. \protect\subref{graph:assembly_all}~Assembly of the Cesic interposer on top of the SDD. \protect\subref{graph:assembly_bonding}~Wire bonding of the SDD to the rigid-flex PCB. \protect\subref{graph:assembly_final}~Detector module after gluing and bonding. \protect\subref{graph:assembly_schematic}~Cross-section of all parts involved in the assembly of the detector module.} 
\label{fig:assembly}
\end{figure*}

\subsubsection*{PEEK mounting jig}
The SDD chip is very fragile, particularly on the entrance window side. For the detector assembly, a special mounting frame made from the thermoplastic polymer PEEK was designed. It supports the detector chip during the mounting procedure without touching any part of the entrance window. 

\subsubsection*{SDD gluing}
One of the most critical steps is the gluing of the SDD chip on the Cesic interposer. Cesic is a material that matches the coefficient of thermal expansion of the SDD and provides a solid mechanical interface to the copper cooling structure. Prior to the gluing, all parts involved in the assembly procedure are cleaned thoroughly. First, the two rigid flex cables are glued on the edges of the Cesic interposer using a two component, electrically insulating, thermally conductive epoxy~(EPO-TEK-920FL~\cite{Epotec_920Fl}). After curing, small stripes of polyethylene terephthalate~(PET) tape~(thickness of~\SI{190}{\micro\meter}) are attached to the outer edges of the large polished surface of the interposer. These PET stripes are used to ensure a defined glue thickness\footnote{In the first iteration of the procedure, a glue containing spherical spacer balls with a diameter of \SI{70}{\micro\meter} was used to create a well defined spacing between the Cesic interposer and the detector. However, it turned out that the spacer balls damaged several detector pixels. As a result, the glue has been replaced and the PET stripes have been included in the assembly to better define the spacing.}. A thin layer of glue with a thickness comparable to the PET stripes is applied to the planar surface of the Cesic interposer as shown in figure~\ref{graph:assembly_cesic}. Afterwards, the SDD chip is placed inside the PEEK jig with the entrance window facing downwards, see figure~\ref{graph:assembly_sdd}. Finally, using an aligner, the Cesic interposer is positioned on top of the SDD chip as illustrated in figures~\ref{graph:assembly_all} and~\ref{graph:assembly_schematic}. Curing is performed in an oven under nitrogen atmosphere at a temperature of~\SI{60}{\celsius} for about~\SI{5.5}{\hour}. 

\subsubsection*{Wire bonding}
In the next step, the detector is electrically connected to the readout electronics using ultrasonic wire bonding. In particular, 361 aluminum wire bonds with a diameter of~\SI{25}{\micro\meter} connect the readout side of the SDD chip with the two rigid flex cables, see figure~\ref{graph:assembly_bonding}. Two additional bond wires from the rigid-flex PCB to the entrance window side are required for the depletion voltages of the SDD chip. To this end, one small corner of the SDD chip is cut away as can be seen in the photograph in figure~\ref{graph:assembly_final}.

\subsubsection*{Electronics assembly}
In the last step, the Cesic interposer holding the SDD is screwed onto the copper cooling block. Thereupon, the two electronics PCBs, called ASIC boards, are mounted on each side of the copper block. The rigid flex cables are connected to the ASIC boards via dedicated Z-ray compression connectors. Finally, the ASIC boards are connected to the bias system and the data acquisition~(DAQ) system using dedicated cables and feedthrough flanges. 

\section{Experimental setups}\label{ch:exp_setups}
Three experimental setups were used to characterize the TRISTAN detector module: 1)~a bench test setup for the characterization with x-rays, 2)~a vacuum chamber with an electron gun for the characterization with electrons, and 3)~the KATRIN monitor spectro\-meter. These setups will be discussed in more detail in the following sections.
\\\\
In all setups, data has been acquired using three synchronized CAEN~VX2740B digitizers. Each unit performs a full waveform digitization for up to 64~channels with a sampling rate of~\SI{125}{\mega\hertz} at a 16-bit resolution. A trapezoidal filter with an exponential baseline correction is applied to reconstruct the energy of the individual events~\cite{JORDANOV1994337}. If not stated otherwise, for the measurements performed in the scope of this work, an energy filter rise time of~$t_{\mathrm{rise}}=\SI{2}{\micro\second}$ and a flat top time of~$t_{\mathrm{top}}=\SI{0.3}{\micro \second}$ have been used. 
\\\\
Most of the measurements presented in the following have been performed with a TRISTAN detector module~(internal nomenclature: S0-166-4) in the bench test setup and in the monitor spectrometer. To validate some of the results, a second detector module~(S0-166-6) has been used for investigations in the electron gun vacuum chamber. The detectors used in this work are prototypes of the final module design. 

\subsection{X-ray bench test setup}\label{ch:exp_setup_mpp}
\begin{figure}[!t]
\centering
\includegraphics[width=1.0\linewidth]{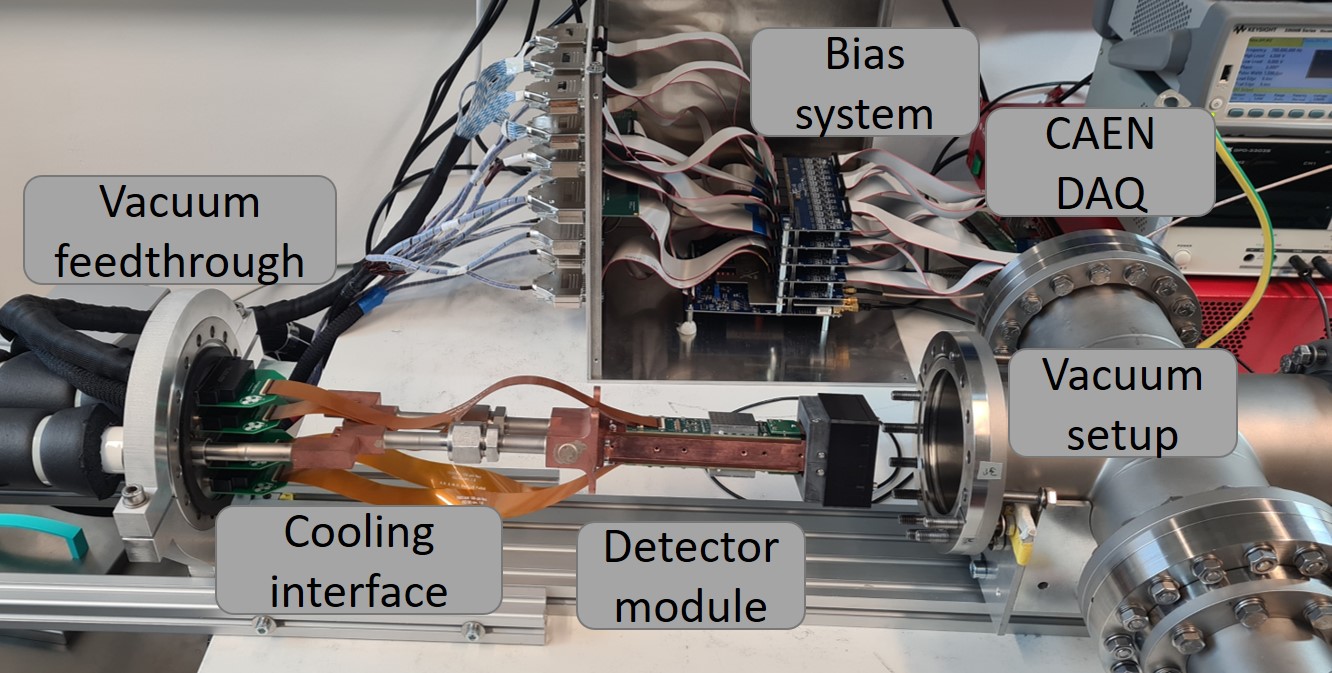}
\caption{Photograph of the X-ray bench test setup. The detector module in the center of the photograph is protected by a cover which at the same time serves as a holding structure for calibration sources. The detector module and the readout electronics are connected to a cooling interface system. The test setup features a feedthrough for all electrical connections as well as for the cooling pipes. Inside the vacuum, the detector and readout electronics are connected via four Kapton flat cables. Outside the vacuum, the signal lines are connected to the bias system via twisted cable pairs. The connection of the signal lines to the DAQ system is realized using short flat cables.}
\label{fig:setup_mpp}
\end{figure}

\begin{figure}[!t]
\centering
\includegraphics[width=1\linewidth]{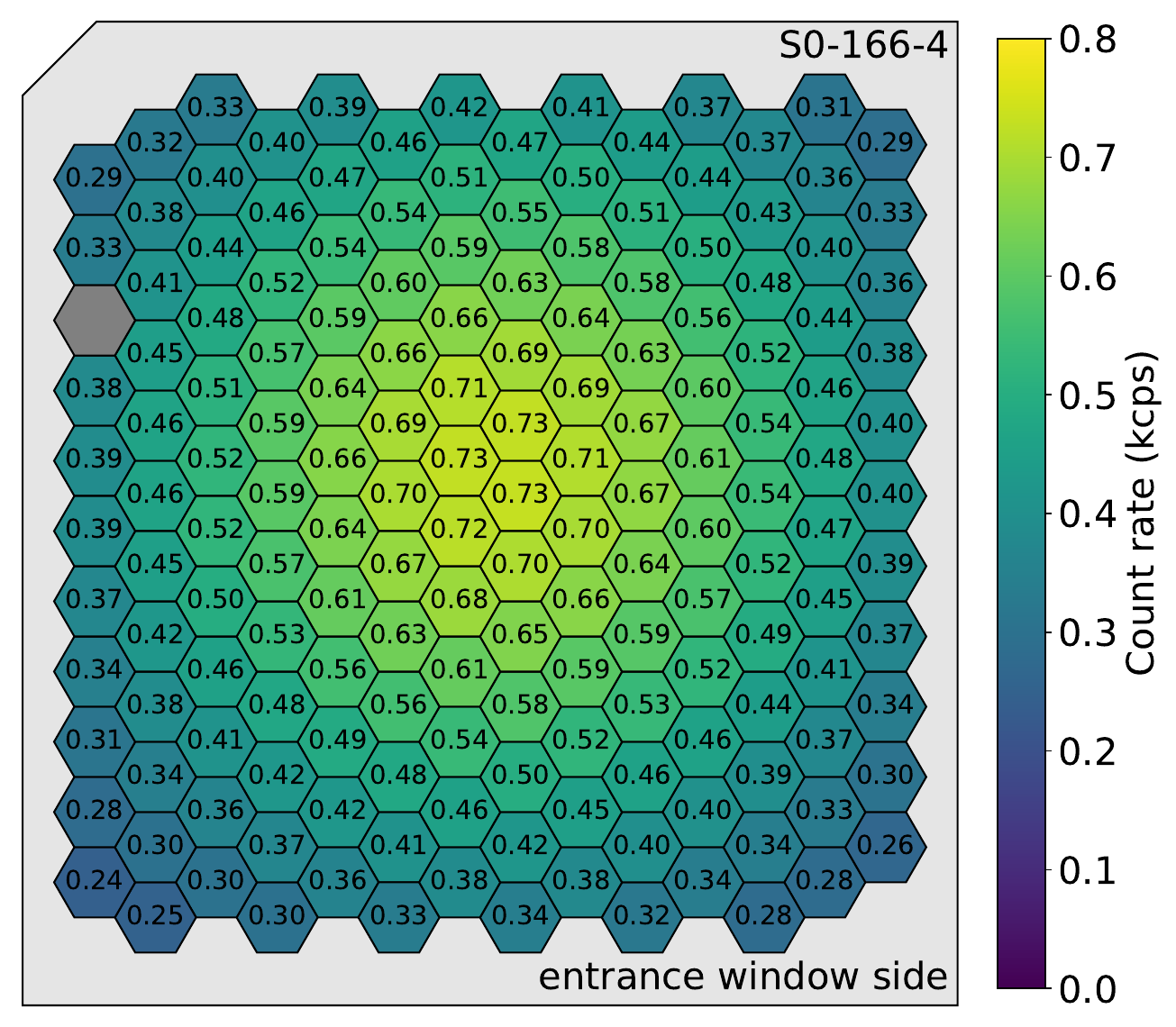}
\caption{Typical pixel map showing the rate distribution measured with an $^{55}$Fe~calibration source in the x-ray bench test setup. The measured count rate for events with energies above~\SI{1.5}{\kilo\electronvolt} is represented by the color bar. The pixels are shown as seen from the entrance window side. The pixel marked in grey had to be disabled due to a loose contact in the signal line.}
\label{fig:pixelmap_countrate_fe55_mpp}
\end{figure}

In a first step, the TRISTAN detector module has been characterized with x-rays in a bench test setup at the Max Planck Institute for Physics, see figure~\ref{fig:setup_mpp}. It consists of a vacuum- and light-tight tube-like vessel made from stainless steel and roughly resembles the geometry of the detector section in the KATRIN monitor spectrometer. The setup is assembled on rails such that the detector module can be accessed easily. The copper block on which the detector module is mounted is connected to a cooling system to operate it at a temperature of~\SI{-35}{\celsius}. A protective cover can be installed at the detector entrance window side to protect the sensitive area of the SDD and the wire bonds. At the same time, this cover can be used for the installation of a radioactive calibration source at a distance of about~\SI{12}{\milli \meter} to the SDD chip. For the detector module characterization with x-rays, $^{55}$Fe and $^{241}$Am sources were used. While the $^{55}$Fe source features two prominent x-ray lines at energies of~\SI{5.9}{\kilo\electronvolt}~($\text{Mn-K}_{\alpha}$) and~\SI{6.4}{\kilo \electronvolt}~($\text{Mn-K}_{\beta}$), respectively, the $^{241}$Am source has several monoenergetic x-ray, gamma ray and, fluorescence lines in the energy range~\SIrange{5}{60}{\kilo\electronvolt}~\cite{LaboratoireNationalHenriBecqerelb_Am241,Spreng_2020}. Representative for the measurements in this work, figure~\ref{fig:pixelmap_countrate_fe55_mpp} shows a typical pixel map of the measured count rate for $^{55}$Fe events with energies above~\SI{1.5}{\kilo\electronvolt}.
%
%

\subsection{Electron gun vacuum chamber}\label{ch:exp_setup_chamber}
\begin{figure}[!t]
\centering
\includegraphics[width=1.0\linewidth]{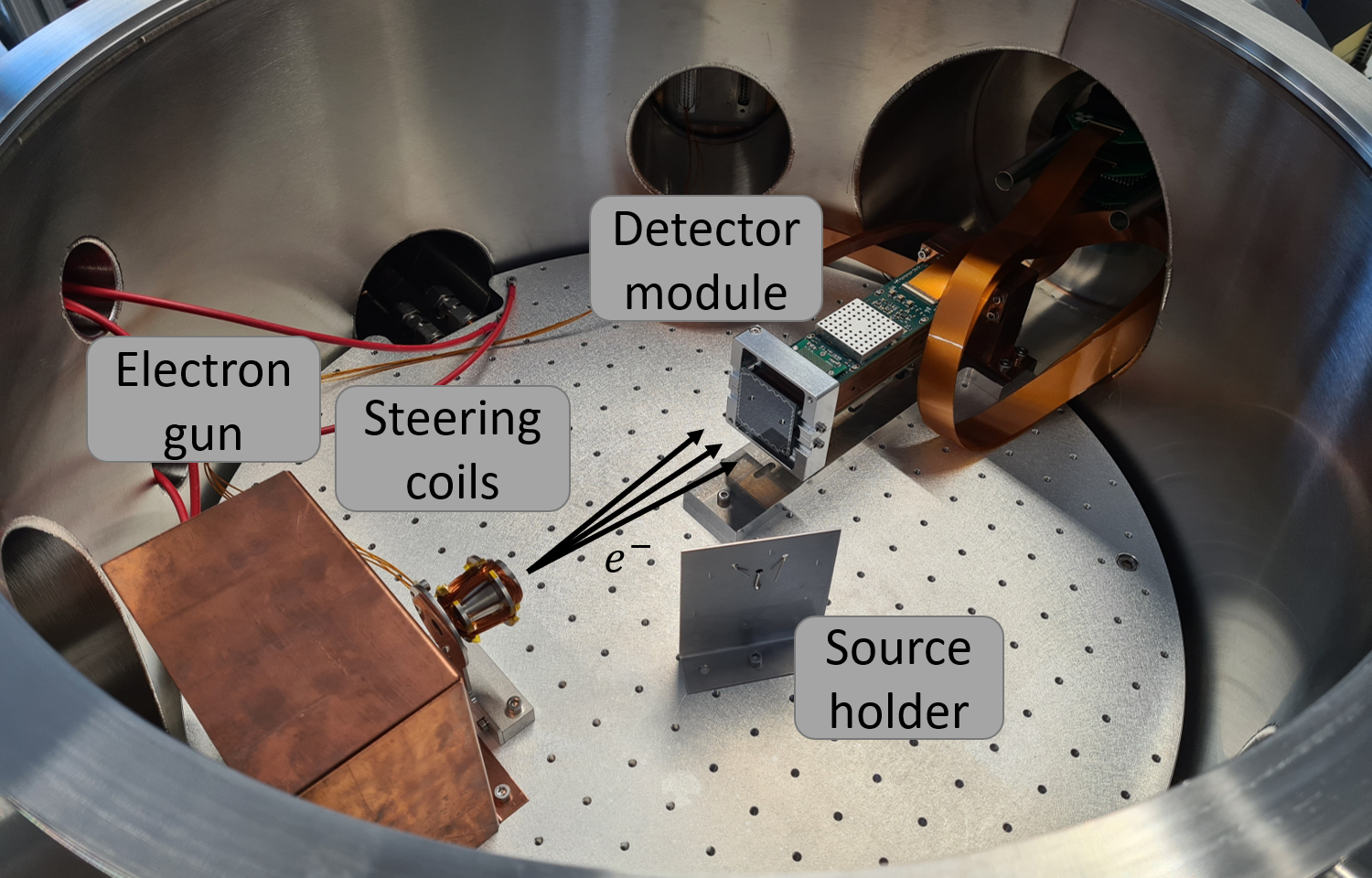}
\caption{Photograph showing the electron gun vacuum chamber. The electron gun is based on a hot cathode using a tantalum wire. Electrons are accelerated via an electric field to energies up to~\SI{20}{\kilo\electronvolt}. They are guided via steering coils towards the entrance window of the detector. The detector module is mounted at a distance of about~\SI{25}{\centi\meter} to the electron source. Parts of the protection cover made from aluminium are still installed on the detector module. Several feedthroughs are used for cable routing, vacuum, and cooling.}
\label{fig:big_vac_chamber}
\end{figure}

The characterization of the detector module with electrons was performed using a customized electron gun installed in a cylindrical vacuum chamber~\cite{Urban_2023}, see figure~\ref{fig:big_vac_chamber}. For the measurements with electrons, good vacuum conditions are required since they have a small free streaming length in air. Using a turbo-molecular pump directly connected to the vacuum chamber, pressure levels as low as~$10^{-7}\,\mathrm{mbar}$ are achieved in the test setup. 
\\\\
The electron gun is based on a hot cathode made from a tantalum filament~(wire diameter of~\SI{25}{\micro\meter}). By applying a voltage and thus heating the wire, electrons are emitted as soon as their thermal energy exceeds \SI{4.3}{\electronvolt} (work function for tantalum)~\cite{JAECKEL_1963}. The filament is positioned such that there is no direct line of sight to the detector. This is necessary to minimize the leakage current caused by the emission of photons from the hot cathode. The electron rate can be set in the range~\SIrange{1}{30}{\kilo\countspersecond} by adjusting the wire temperature. The tantalum wire is mounted in a stainless steel cage which is connected to a high voltage power supply. By applying an electric field of up to~\SI{20}{\kilo\volt} to the cage, electrons are accelerated towards the detector module. The beam illuminates roughly \num{10}~pixels at once. Magnetic steering coils mounted in front of the electron gun are used to guide the electrons such that the entire entrance window surface of the detector is illuminated homogeneously~\cite{Urban_2023}.

\subsection{KATRIN monitor spectrometer}\label{ch:exp_setup_mos}
\begin{figure}[!t]
\centering
\includegraphics[width=1.0\linewidth]{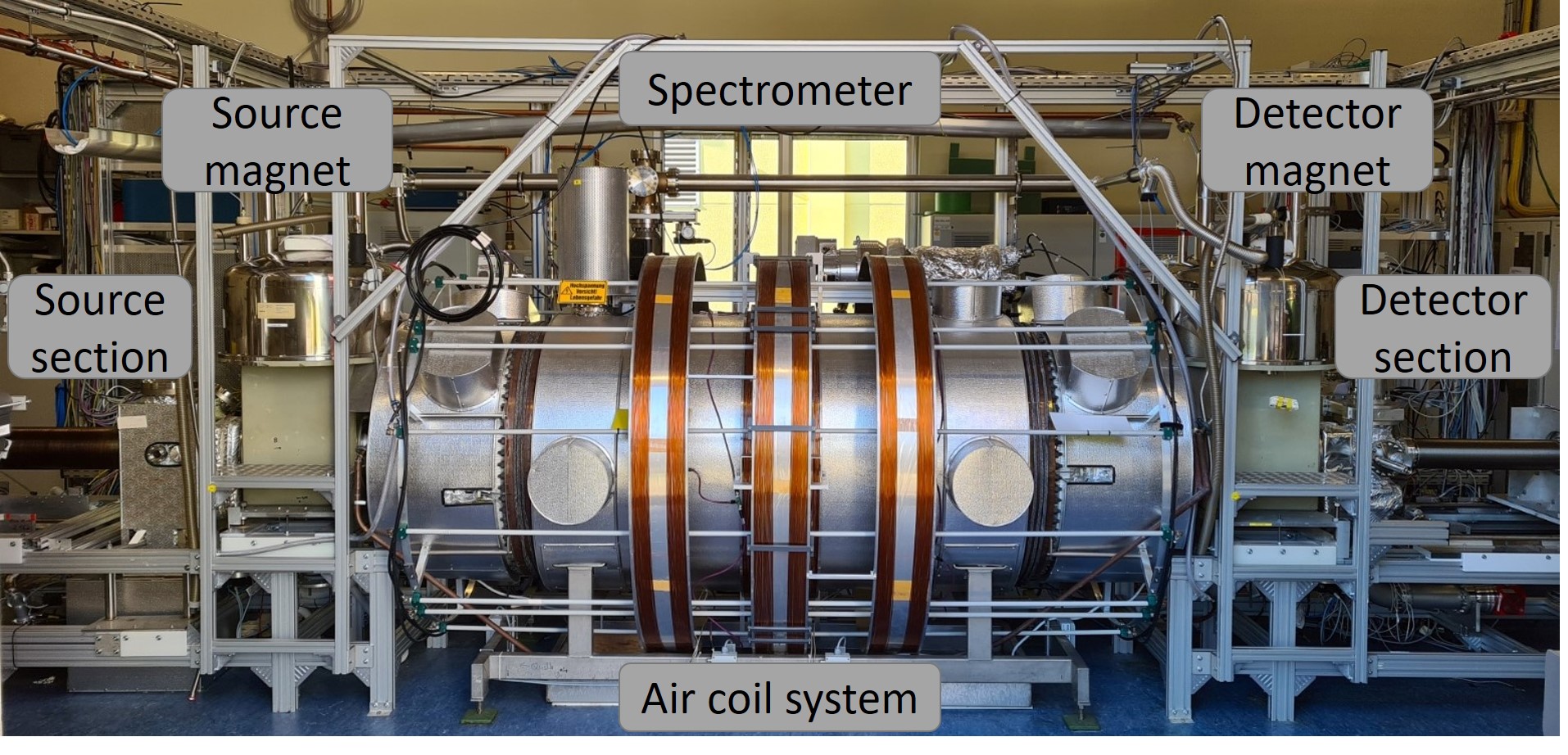}
\caption{Photograph of the KATRIN monitor spectrometer. The spectrometer is shown in the center, while the two superconducting magnets can be seen on each side. The spectrometer is surrounded by three copper coils which shape the magnetic field inside the vessel. The evaporated $^{\mathrm{83m}}$Kr~source is installed in the source section at a position close to the source magnet. The TRISTAN detector module is mounted in the detector section close to the detector magnet.}
\label{fig:Monitorspectrometer}
\end{figure}

In order to characterize the TRISTAN detector module in a more realistic environment~(magnetic field, vacuum, high voltage, etc.), it has been installed in the KATRIN monitor spectrometer. A photograph of the apparatus is shown in figure~\ref{fig:Monitorspectrometer}. The monitor spectrometer was initially used as a high voltage monitoring device for the KATRIN neutrino mass measurements. More recently, it has been repurposed as a test facility for potential hardware upgrades of the KATRIN experiment. The spectrometer as well as many other parts come from the former Mainz neutrino mass experiment, one of the predecessors of KATRIN~\cite{Bonn_2002}. The spectrometer is a stainless vessel with a diameter of~\SI{1}{\meter} and a length of about~\SI{3}{\meter}. Two superconducting magnets provide a strong magnetic field of up to~\SI{6}{\tesla} on each side of the spectrometer~\cite{PHDSlezak_2015}. The spectrometer vessel is surrounded by a system of air coils which shapes the magnetic field and compensates for the earth magnetic field. A set of cylindrical and conical electrodes mounted inside the spectrometer allows fine adjustment of the electric fields.
%
\\\\
The operating principle of the monitor spectrometer is based on the MAC-E filter principle which is also utilized in the KATRIN experiment~\mbox{\cite{PICARD1992345,Angrik_2005,Aker_2021}}. Electrons created in a high magnetic field~(\SI{3.6}{\tesla}) in the source section are guided on a cyclotron motion along the magnetic field lines towards the spectrometer. Due to the reduction of the magnetic field strength towards the center of the spectrometer~(\SI{3}{\milli\tesla}), the transverse momentum of the electrons is adiabatically transformed into a longitudinal one. By applying a retarding voltage to the electrodes of the spectrometer, an electric field parallel to the magnetic field lines is created. This field acts as a high-pass filter to the longitudinal component of the electron momentum. All electrons with a longitudinal momentum sufficient to overcome the retarding potential are magnetically guided to the TRISTAN detector in the detector section. At the position of the detector, the magnetic field has a value of~\SI{100}{\milli \tesla}. In addition, a vacuum level of below~\SI{e-8}{\milli bar} at the detector position and \SI{e-10}{\milli bar} in the spectrometer can be achieved.
\\\\
To characterize the TRISTAN detector in the monitor spectrometer, three different types of x-ray and electron sources were used.

\subsubsection*{\(^{55}\)Fe source}
To investigate the detector system with x-rays, an $^{55}$Fe calibration source was installed in the beamline of the monitor spectrometer inside the detector section. The source was moved in front of the detector using a movable source holder~\cite{Bruch_2023}. 

\subsubsection*{\(^{\mathrm{83m}}\)Kr source}
The detector response to electrons was characterized using an $^{83\mathrm{m}}$Kr source evaporated on the surface of a highly oriented pyrolithic graphite (HOPG) substrate. It is installed in front of the superconducting solenoid on the source section side of the spectrometer. As listed in table~\ref{tab:Krypton}, it provides quasi-monoenergetic conversion electrons in the energy range from~\SIrange{7}{32}{\kilo \electronvolt}~\mbox{\cite{Venos_2018, Picard_1992}}. In the scope of this work, internal conversion electrons of the gamma decay with an energy of~\SI{32.15}{\kilo\electronvolt} were used exclusively. For the energy calibration, a weighted average of corresponding sublines was calculated. This is required since the detector cannot resolve energetic separations on the eV scale. As an example, the L-32 electron line corresponds to the weighted average of the L$_1$, L$_2$, and L$_3$ lines. The electron energy~$E_{\text{Kr}}$ as measured with the detector for a given electron line energy~$E_{\mathrm{line}}$ can be described by the equation
\begin{equation}
E_{\text{Kr}} = E_{\mathrm{line}} + \left|e\cdot U_{\text{sc}}\right| - |e\cdot U_{\mathrm{bias}}|.\label{eq:electron_kr}
\end{equation}
Here, $e$ denotes the elementary charge, and $U_{\text{sc}}$ an additional source potential which was set to a value of~\SI{-500}{\volt} to effectively increase the electron emission rate towards the detector. The quantity~$U_{\mathrm{bias}}=\SI{-115}{\volt}$ indicates the biasing potential on the entrance window side of the detector.
\begin{table}  
\centering
\begin{tabular}{ c c c c}
\multirow{2}{*}{Line} & \multirow{2}{*}{Energy (\SI{}{\electronvolt})} & Intensity & Weighted \\ 
&  & per decay (\SI{}{\percent}) & average (\SI{}{\electronvolt})\\ 
 \toprule
$K$ & $17824.2(5)$ & $24.8(5)$ & \\
\hline
$L_1$ & $30226.8(9)$ & $1.56(2)$ & \\
$L_2$ & $30419.5(5)$ & $24.3(3)$ & \\
$L_3$ & $30472.2(5)$ & $37.8(5)$ & \multirow{-3}{*}{30446.1}\\
\hline
$M_1$ & $31858.7(6)$ & $0.249(4)$ & \\
$M_2$ & $31929.3(5)$ & $4.02(6)$ & \\
$M_3$ & $31936.9(5)$ & $6.24(9)$ & \\
$M_4$ & $32056.4(5)$ & $0.0628(9)$ & \\
$M_5$ & $32057.6(5)$ & $0.0884(12)$ & \multirow{-5}{*}{31933.9}\\
\bottomrule
\end{tabular}
\caption{Conversion electrons from the $^{83\mathrm{m}}$Kr source used for the experimental investigations~(values corresponding to the \SI{32.15}{\kilo\electronvolt} $\gamma$-decay). The weighted average of the $L$ and $M$ lines is calculated using the intensity per decay. Values taken from~\cite{Venos_2018}.}
\label{tab:Krypton}
\end{table}
\noindent
\subsubsection*{Wall electrons}
Secondary electrons emitted from the inner spectrometer electrode surface were used as a second electron source. In the following, we will refer to them as wall electrons. These electrons are created by high-energetic particles such as muons which interact with the spectrometer material facing the vacuum side. The electrons leave the surface with energies of~\mbox{$E_{\mathrm{sec}}<\SI{30}{\electronvolt}$}~(peak energy at~$E_{\mathrm{sec}}=\SIrange{1}{2}{\electronvolt}$)~\cite{ALTENMULLER_2019}. They are then accelerated in the direction of the detector via the potential~$U$ of the electrode system. The electron energy at the detector can be described by the equation
\begin{equation}
E_{\mathrm{wall}} = E_{\mathrm{sec}} + \left|e\cdot U\right| - |e\cdot U_{\mathrm{bias}}|. \label{eq:electron_wall}
\end{equation}
In the standard measurement configuration, the wall electrons are magnetically shielded from the spectrometer volume and are not detected~\cite{Mainz_2005}. In order to use these electrons for calibration and characterization purposes, the magnetic fields in the spectrometer can be intentionally adjusted such that they touch the walls. Consequently, the electrons are guided towards the detector. For the measurements performed in the scope of this work, wall electrons with energies from~\SIrange{10}{32}{\kilo\electronvolt} and a total count rate of~\SI{25}{\countspersecond} were used. 

\section{Characterization of the TRISTAN detector modules}\label{ch:characterization}
In this section, we present the results of the characterization of the TRISTAN detector with x-rays and electrons. In particular, properties such as the spectral response, the energy resolution, and the stability of the system will be discussed.

\subsection{Detector response to x-rays}
\begin{figure*}[!t]
\subfloat[ $^{55}$Fe spectrum\label{fig:energy_spectrum_fe55_mpp}]
{\includegraphics[width=0.30\textwidth]{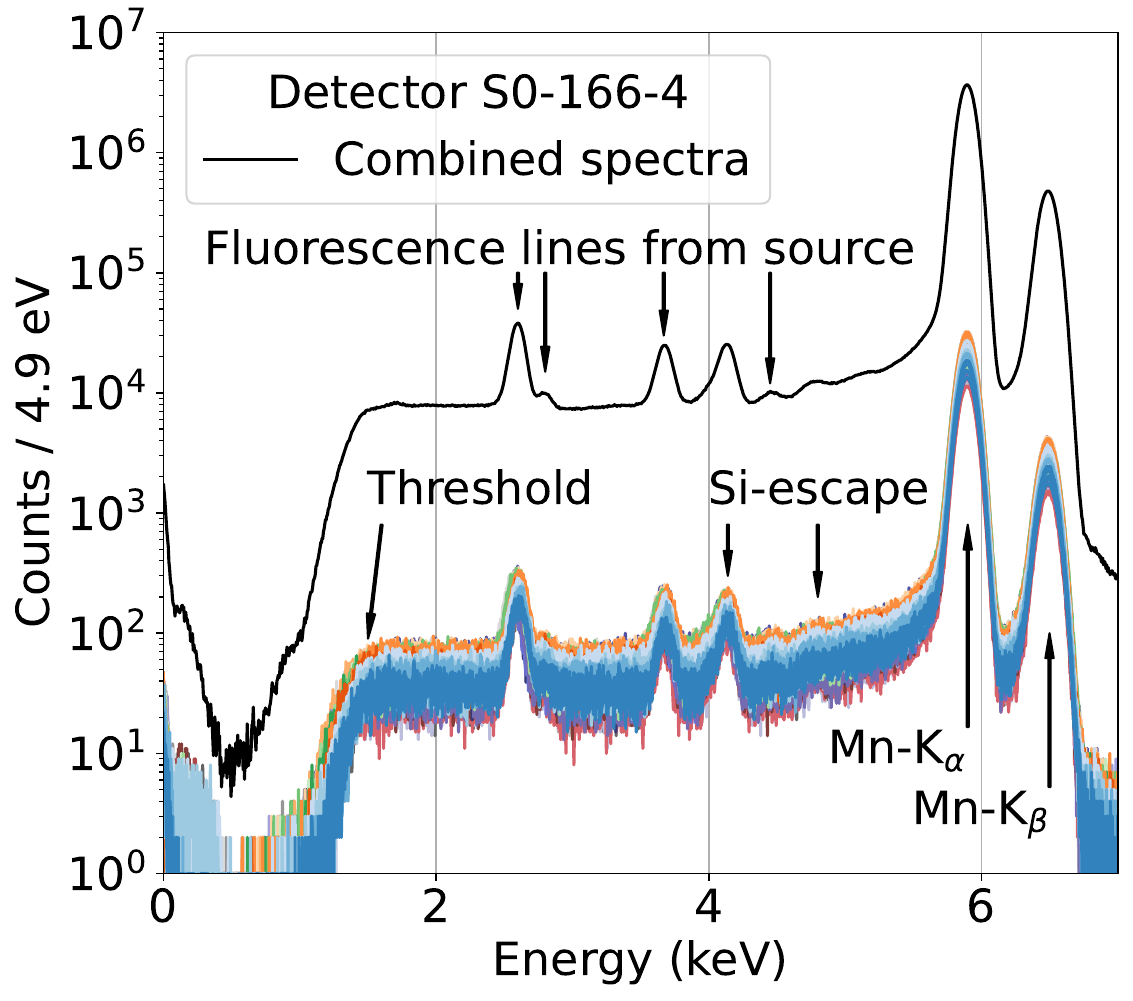}
}\quad
\subfloat[\SI{20}{\kilo\electronvolt} electrons\label{fig:energy_spectrum_20keV_tum}]
{\includegraphics[width=0.30\textwidth]{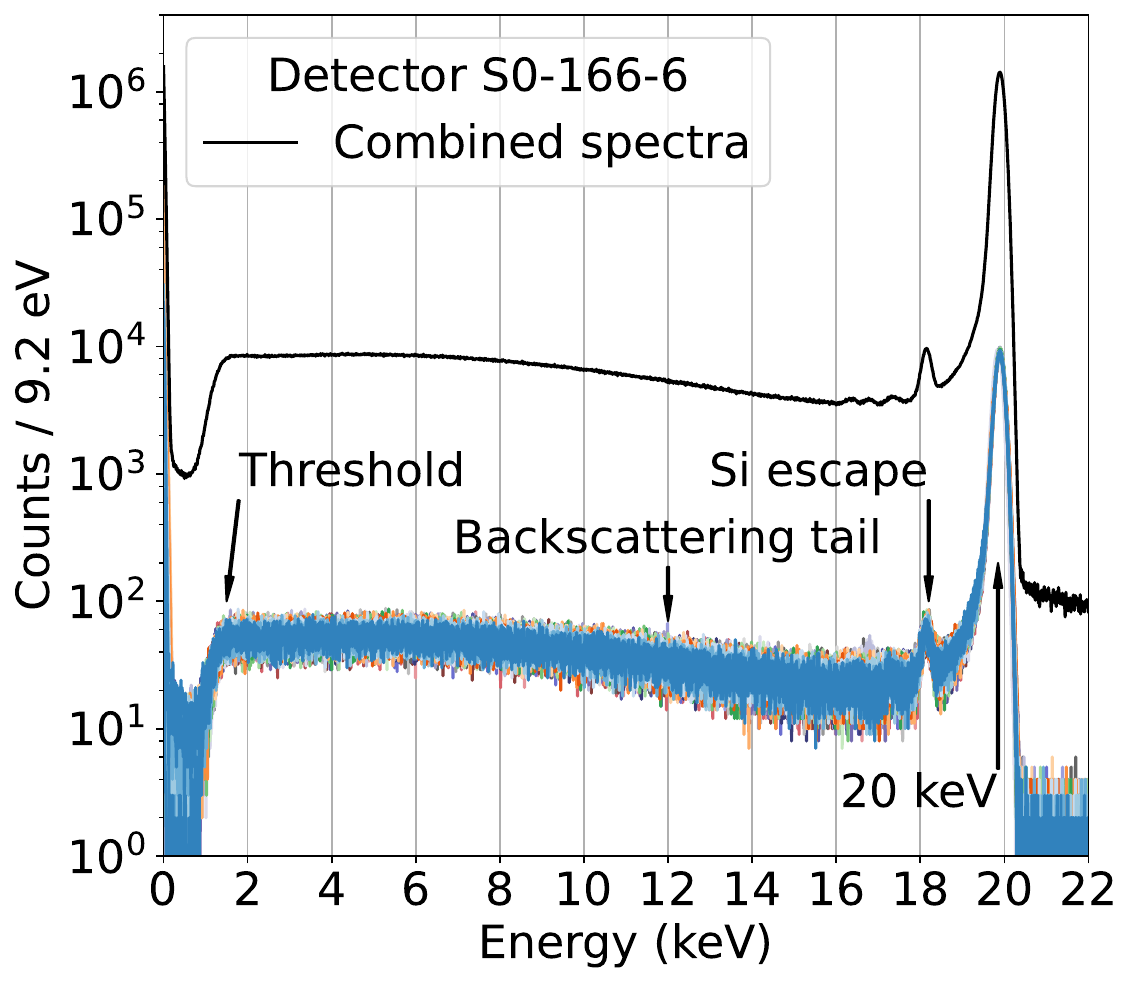}
}\quad
\subfloat[L-32 and M-32 lines of $^{\mathrm{83m}}$Kr\label{fig:spectrum_kr83m}]
{\includegraphics[width=0.359\textwidth]{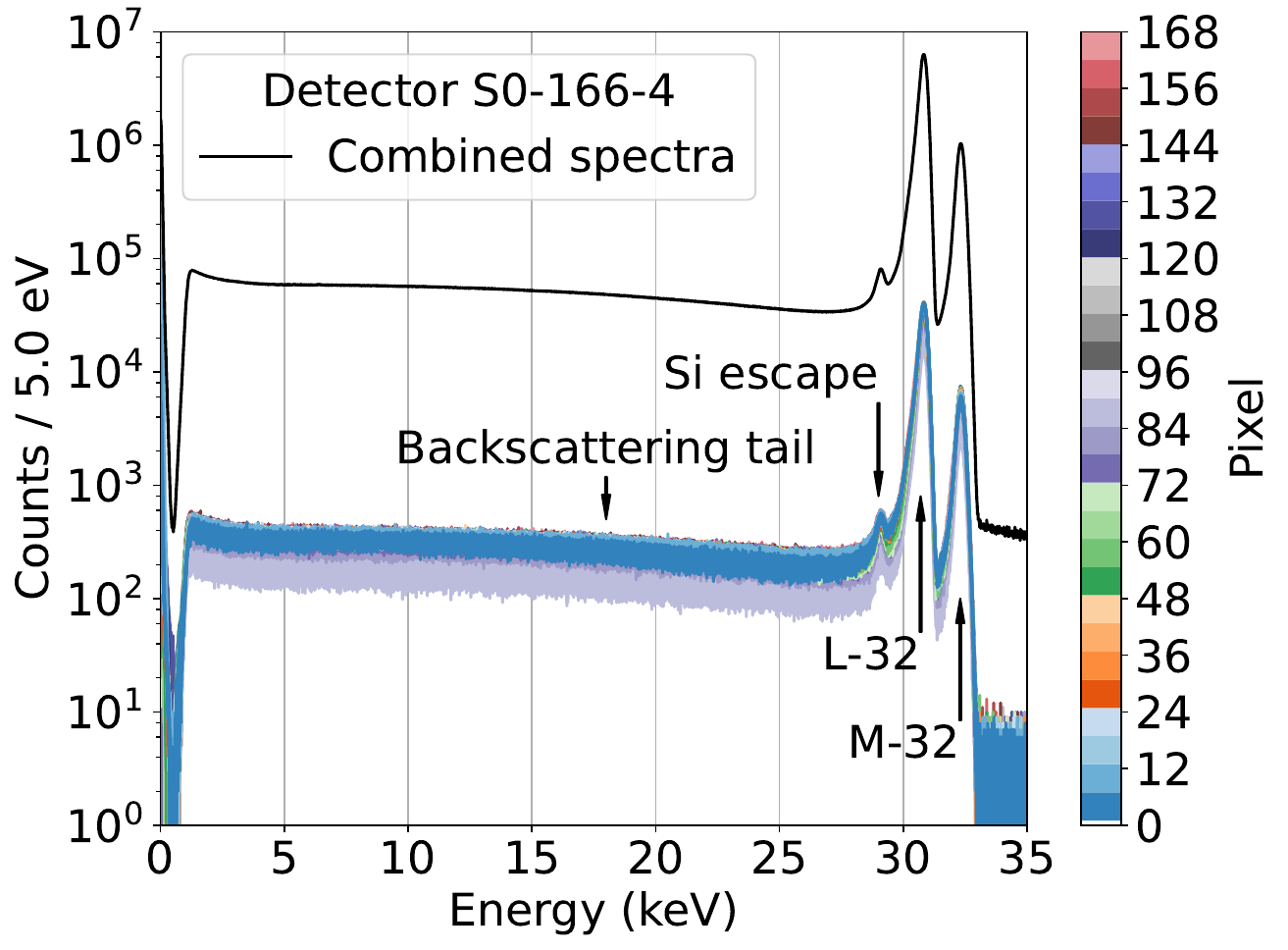}}\quad 
\caption{Recorded energy spectra in the presence of x-ray and electron sources in the three different experimental setups. The black curve corresponds to the superposition of all energy spectra. The color bar indicates the individual pixels. \protect\subref{fig:energy_spectrum_fe55_mpp}~Measurement with an $^{55}$Fe calibration source placed in front of the detector in the x-ray bench test setup. \protect\subref{fig:energy_spectrum_20keV_tum}~Electron energy spectra measured with an electron gun at \SI{20}{\kilo\electronvolt} in the vacuum chamber setup. In the combined spectrum, a small non-physical distortion can be observed at about~\SI{17}{\kilo\electronvolt}. It originates from an instability of the electron gun and is currently under investigation. \protect\subref{fig:spectrum_kr83m}~Energy spectra of the \mbox{L-32} and \mbox{M-32} lines in the presence of a $^{\text{83m}}$Kr source in the monitor spectrometer. The source potential was set to $U_{\mathrm{sc}}=\SI{-500}{\volt}$ and the retarding potential to $U=\SI{30.2}{\kilo\electronvolt}$.} 
\label{fig:spectrum}
\end{figure*}

For the characterization of the detector response to x-rays, an $^{55}$Fe calibration source was used. A typical energy spectrum recorded in the x-ray bench test setup is shown in figure~\ref{fig:energy_spectrum_fe55_mpp}. The $\text{Mn-K}_{\alpha}$ and $\text{Mn-K}_{\beta}$ lines at~\SI{5.9}{\kilo \electronvolt} and \SI{6.5}{\kilo \electronvolt}, respectively, are clearly separated with an average energy resolution of \SI{143.7}{\electronvolt}~FWHM and \SI{151.7}{\electronvolt}~FWHM, respectively. Here, the energy resolution is extracted for each pixel and then averaged. In addition, the silicon escape peak corresponding to the~$\text{Mn-K}_{\alpha}$ line can be seen. Here, the incident radiation ionizes an electron of the K-shell of the detector material. The created hole is filled by an electron from a higher-lying shell, emitting a photon with an energy of~\SI{1.74}{\kilo\electronvolt}~\cite{thompson_2001}. This photon can be either reabsorbed by the detector or leave it undetected, leading to the silicon escape peak. At lower energies, various fluorescence lines inherent to the radioactive source are visible in the spectrum. 
\\\\
For energies above~\SI{1}{\kilo \electronvolt}, the absorption length of x-rays in silicon is~$\geq\SI{1}{\micro\meter}$~\cite{NIST} and therefore much larger than the thickness of the detector entrance window with values on the order of~\mbox{\SI{50}{\nano\meter} \cite{Mertens_2021}}. Therefore, most of the photons deposit their entire energy in the fully sensitive detection volume. As a result, the detector response is only slightly affected by entrance window effects and the $\text{Mn-K}_{\alpha}$ and $\text{Mn-K}_{\beta}$ lines can be approximated by a Gaussian distribution. A more detailed description on the spectral response to x-rays can be found in~\cite{Eggert_2004}. 
%
\subsection{Detector response to electrons}
To investigate the detector response to electrons, the detector modules have been illuminated with monoenergetic electrons with energies in the range from~\SIrange{5}{20}{\kilo\electronvolt} in different setups. A typical electron energy spectrum~(recorded with the detector S0-166-6) is shown in figure~\ref{fig:energy_spectrum_20keV_tum}. The measurement was performed using the electron gun in the vacuum chamber setup. Compared to the x-ray spectra, several distinct features can be observed in the spectral response. 

\subsubsection*{Peak shift and low-energy tail}
As opposed to photons, electrons interact much more strongly with the detector material due to their massive and charged nature. This results in an energy deposition profile with a pronounced component in close proximity to the entrance window surface. Here, the electric field is not strong enough to transport all charge carriers to the readout anode. This reduces the charge collection efficiency which is therefore a function of the interaction depth~\cite{Eggert_2004}. Since a fraction of the energy is lost in the region of the entrance window, the measured energy is shifted to lower values. In addition, the full energy peak becomes asymmetric featuring a pronounced low-energy tail. This effectively decreases the energy resolution of the detection system. The effect of the entrance window on the detector response decreases with higher electron energies and smaller entrance window thicknesses, since then a smaller fraction of the total energy is lost. For previous TRISTAN detectors consisting of 7~pixels, the energy shift has been determined to be~\SI{29(2)}{\electronvolt} for electrons with an energy of~\SI{30.4}{\kilo\electronvolt}~(using the M-line of a $^{\text{83m}}$Kr source). Assuming a simplified step-like dead layer model for the entrance window charge collection efficiency, an entrance window thickness of~\SI{49(3)}{\nano\meter} was determined~\cite{Mertens_2021}. This value can only be considered as a rough estimation; a more realistic model can be found in~\cite{Nava_2021}.

\subsubsection*{Backscattering tail}
Another effect of incomplete charge collection is due to the backscattering of electrons. The incoming electrons as well as the secondary electrons undergo scattering inside the detector material which changes their momenta. As a result, the electrons can change their directions such that they eventually leave the detector before having deposited their entire energy in the sensitive volume. In the energy spectrum, this gives rise to a backscattering tail, i.e.~an extended continuum towards lower energies. For low-energetic electrons in the~\SI{}{\kilo\electronvolt} range, the probability of backscattering is on the order of~\SI{20}{\percent}~\cite{Darlington_1972}. The amount of backscattering decreases with higher electron energies and with smaller incidence angles (the incident angle of an electron beam perpendicular to the detector surface is~\SI{0}{\degree}).

\subsubsection*{Electron response in the monitor spectrometer}
To test the electron response of the TRISTAN detector module in an environment that closely resembles the KATRIN experiment, in particular the MAC-E filter, a detector module~(S0-166-4) has been installed in the monitor spectrometer. As described in Section~\ref{ch:exp_setup_mos}, the apparatus provides different quasi-monoenergetic electron sources. Typical energy spectra for the individual pixels measured in the presence of a $^{\text{83m}}$Kr source are shown in figure~\ref{fig:spectrum_kr83m}. To increase the rate of electrons hitting the detector, the source potential was set to~$U_{\mathrm{sc}}= \SI{-500}{\volt}$. In this measurement, the retarding potential was set to a value of~$U=\SI{30.2}{\kilo \volt}$. Therefore, only the two main peaks corresponding to quasi-monoenergetic electrons of the~\mbox{L-32} and \mbox{M-32} lines can be observed. In addition, the silicon escape peak corresponding to the~\mbox{L-32} line is visible, while the one corresponding to the~\mbox{M-32} line is superimposed on the main peak of the~\mbox{L-32} line and thus not visible. At smaller energies, the backscattering continuum can be observed. 
\\\\
The electron response of the detector was fitted with a heuristic model that accounts for the features described above: The region of the main peak is described by a Gaussian distribution in combination with an exponential low-energy tail. In addition, also the backscattering continuum is taken into account using a higher order polynomial. More details on the model are discussed in~\cite{Mertens_2021}.

\subsection{Energy resolution and pixel homogeneity}\label{ch:homogeneity}
In this section, the energy resolution and the homogeneity of the detector response is investigated. Besides the contribution from Fano statistics which is inherent to every semiconductor detector system, the energy resolution of x-rays is mainly determined by the electronic noise of the detector and the readout electronics. As discussed above, in the case of electrons, the energy resolution is also affected by entrance window effects. Particularly at lower energies, these entrance window effects become more relevant. 

\subsubsection*{X-ray energy resolution}
To determine the energy resolution for x-rays, the $\text{Mn-K}_{\alpha}$~peak of the $^{55}$Fe spectrum is approximated with a Gaussian and the full width at half maximum~(FWHM) is extracted. The measured energy resolution of all pixels is shown in the pixel map in~figure~\ref{fig:pixelmap_fwhm_fe55_mpp}. An average energy resolution of~$\SI{143.7}{\electronvolt}$~FWHM (\SI{2}{\micro\second} filter rise time) at the~$\text{Mn-K}_{\alpha}$ peak was obtained~(detector S0-166-4). The energy resolution of all pixels varies by at most~\SI{10}{\electronvolt} which is equivalent to \SI{7}{\percent} when compared to the average energy resolution. This is an indication that the electronic noise of the detector and the readout electronics is comparable for the individual pixels.

\subsubsection*{Entrance window effects}
The detector S0-166-4 has been illuminated with electrons for the first time in the monitor spectrometer. The energy resolution for electrons of the L-32 line measured with a~$^{\text{83m}}$Kr source is shown in figure~\ref{fig:pixelmap_fwhm_l32_mos}. In the pixel map, two distinct areas can be observed: one area has the shape of a semicircle with improved energy resolution values, while the other area exhibits worse resolution values. This behavior can be related to a thicker entrance window in the region with worse energy resolution values. When looking at the energy resolution pixel maps, it is only observed for electrons and not for x-rays. This is due to the fact that x-rays are negligibly affected by a slightly increased entrance window thickness. However, the effect can also be made visible for x-rays by taking the ratio between the measured count rate inside the main region of the $\text{Mn-K}_{\alpha}$ line and the count rate in the lower energy region~(just below the peak), see figure~\ref{fig:pixelmap_photon_ratio}. It shows the same pattern as the electron data in figure~\ref{fig:pixelmap_fwhm_l32_mos}. An explanation of the entrance window effect for x-rays on this region can be found in e.g. \cite{Eggert_2004}.  
\\\\
To pin down the origin of the effectively thicker entrance window, many tests have been performed. The origin could be related to the deposition of an additional water-soluble substance on the entrance window surface during the assembly procedure. This substance effectively increases the thickness of the entrance window. Fortunately, it can be removed using a dedicated cleaning procedure, i.e.~rinsing the detector carefully with acetone, isopropyl alcohol and de-ionized water. This results in a very homogeneous detector performance compatible with the energy resolution values in the semicircle area.  

\subsubsection*{Electron energy resolution}
\begin{figure*}
\subfloat[X-ray resolution map ($^{55}$Fe source, $\text{Mn-K}_{\alpha}$ line)\label{fig:pixelmap_fwhm_fe55_mpp}]{\includegraphics[width=0.49\textwidth]{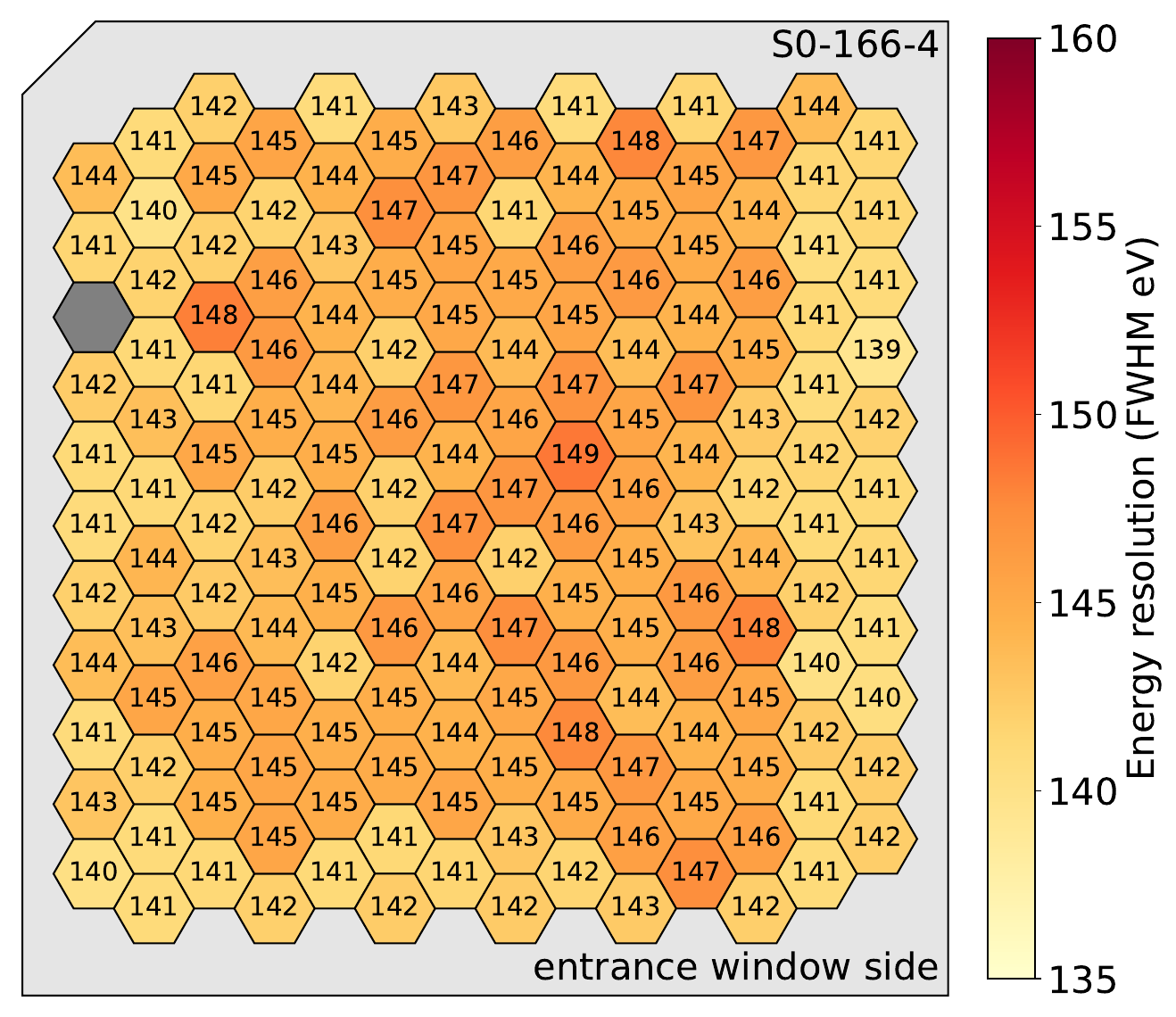}
}\quad
\subfloat[Electron resolution map ($^{\text{83m}}$Kr source, L-32 line)\label{fig:pixelmap_fwhm_l32_mos}]{\includegraphics[width=0.49\textwidth]{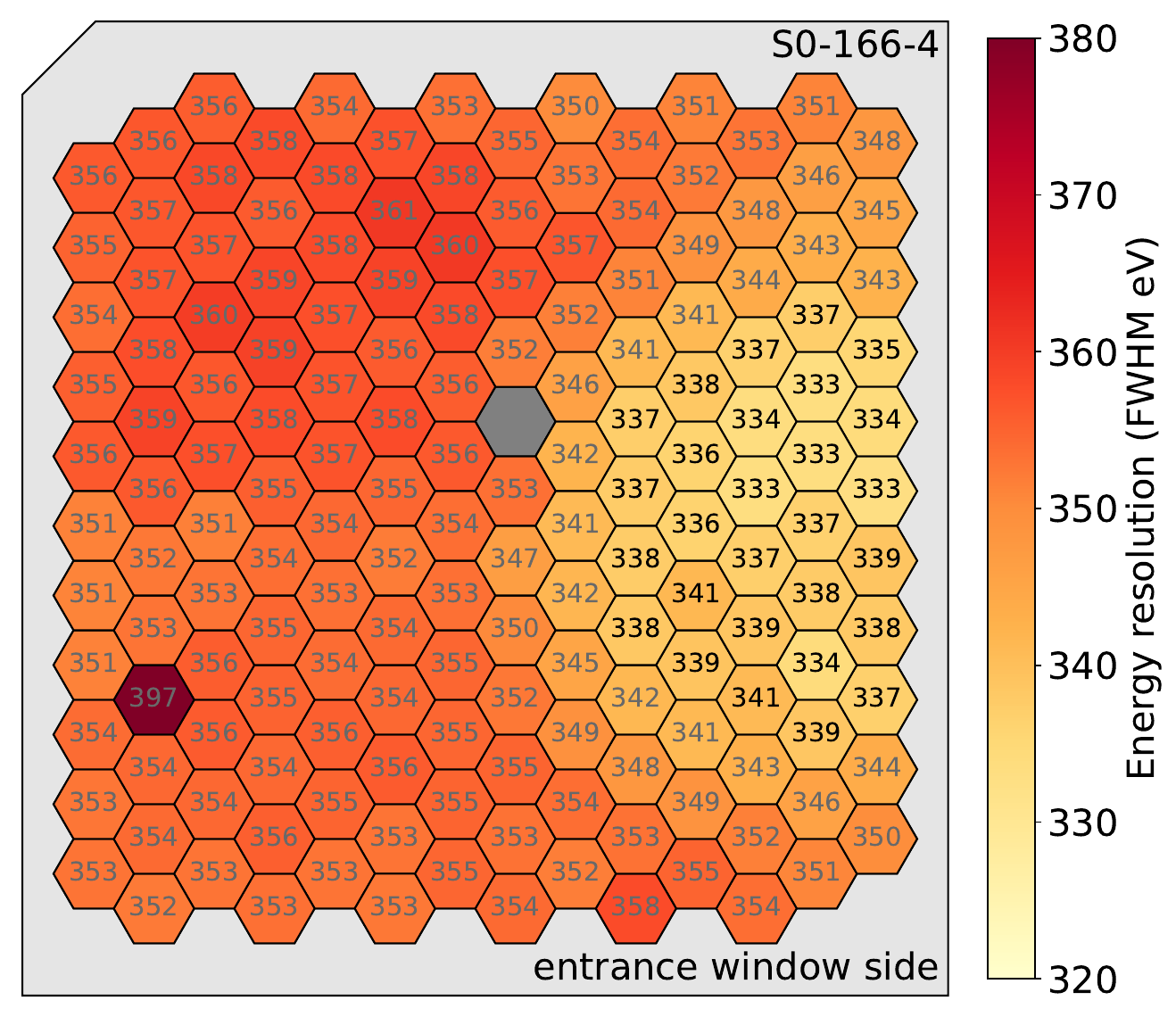}
}\quad
\subfloat[Entrance window effects in x-ray measurement\label{fig:pixelmap_photon_ratio}]{\includegraphics[width=0.49\textwidth]{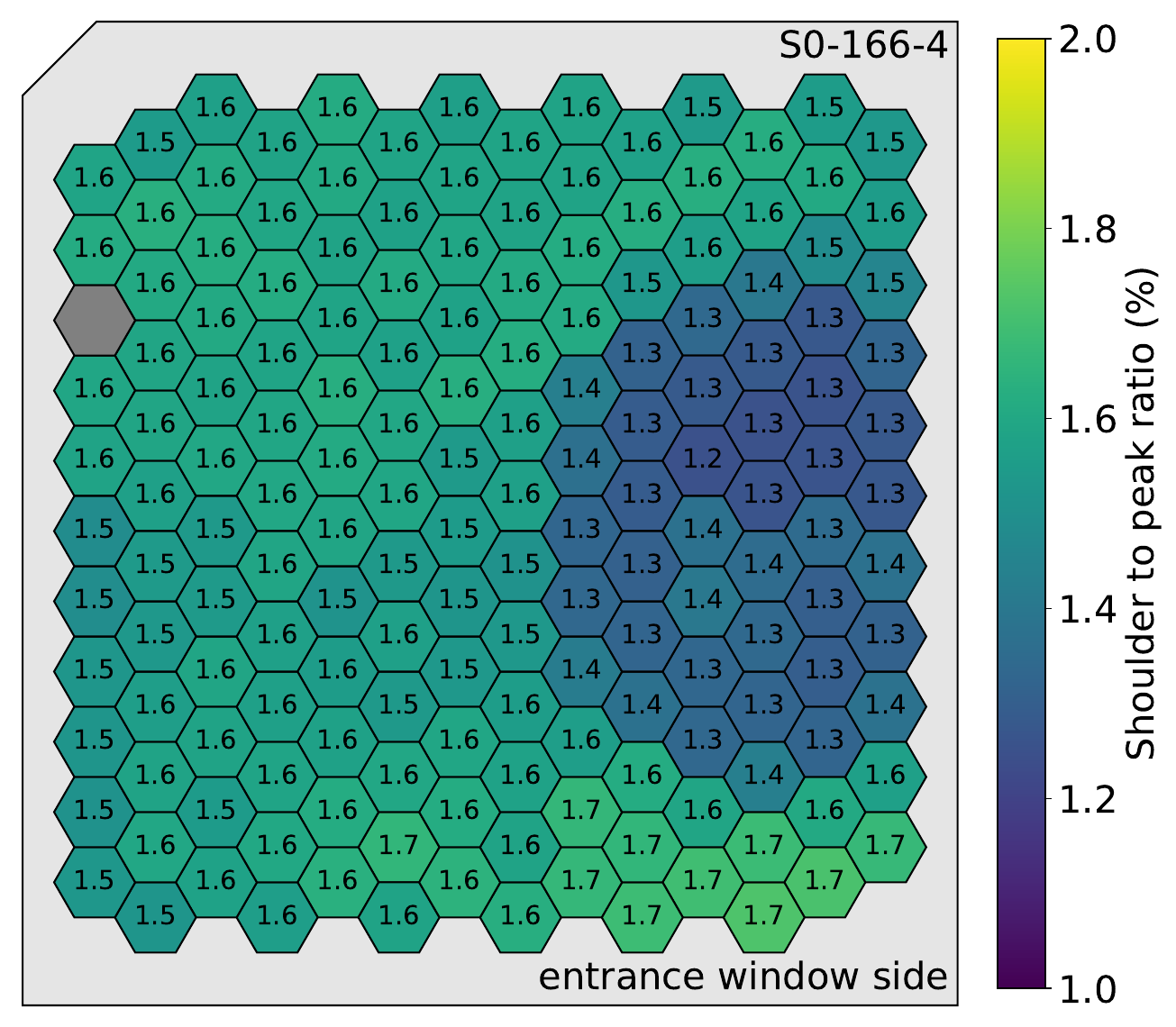}
}\quad
\subfloat[Electron resolution map (\SI{20}{\kilo\electronvolt} electrons, electron gun)\label{fig:pixelmap_fwhm_electron_tum}]{\includegraphics[width=0.49\textwidth]{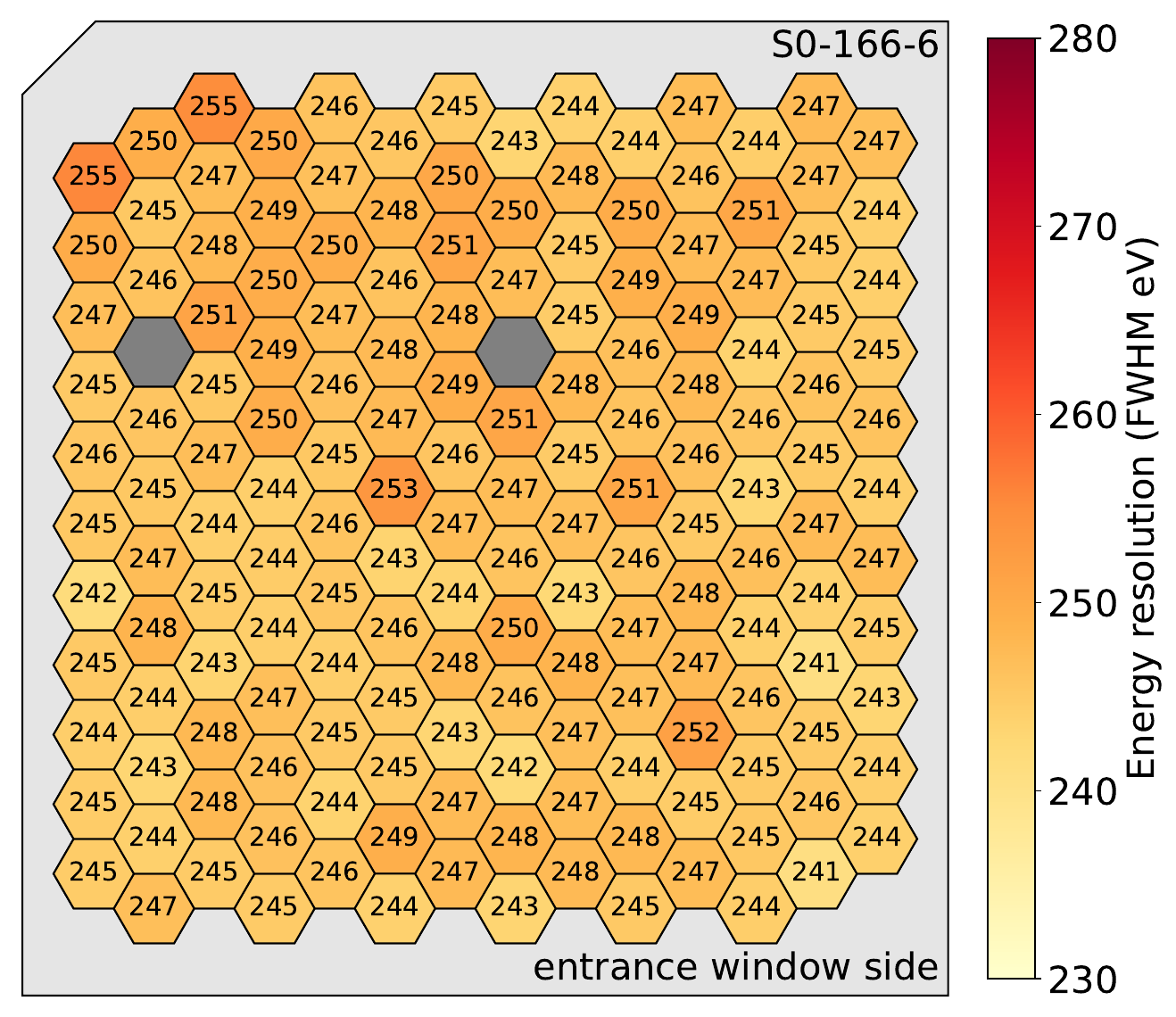}
}\quad
\caption{Pixel maps of the detectors S0-166-4 and S0-166-6 showing the energy resolution and entrance window effects. Disabled pixels are colored in grey. \protect\subref{fig:pixelmap_fwhm_fe55_mpp}~Measured energy resolution of the $\text{Mn-K}_{\alpha}$ line at~\SI{5.9}{\kilo\electronvolt} using an $^{55}$Fe source in the test bench setup. The average energy resolution of all pixels is~$\SI{143.7}{\electronvolt}$~FWHM (energy filter rise time of~\SI{2}{\micro\second}). \protect\subref{fig:pixelmap_fwhm_l32_mos}~Pixel map showing the energy resolution for electrons of the L-32 line at~\SI{30.8}{\kilo\electronvolt} using a $^{\text{83m}}$Kr source in the monitor spectrometer. The average energy resolution of all pixels is~\SI{336.7}{\electronvolt}~FWHM~(energy filter rise time of~\SI{2}{\micro\second}). The pixels with grey values correspond to pixels with an additional substance on the detector that could be removed after the measurement campaign had been completed. \protect\subref{fig:pixelmap_photon_ratio}~Entrance window effects observed in the $^{55}$Fe energy spectra (same measurement as in~\protect\subref{fig:pixelmap_fwhm_fe55_mpp}). The map shows the ratio of the counts in the range from~\SIrange{5.2}{5.6}{\kilo\electronvolt} (region corresponding to entrance window effects) and the counts in the range from~\SIrange{5.8}{6.1}{\kilo\electronvolt} (full energy deposition). It shows the same structure as the energy resolution of electrons in \protect\subref{fig:pixelmap_fwhm_l32_mos}. \protect\subref{fig:pixelmap_fwhm_electron_tum}~Measured energy resolution of monoenergetic electrons~(\SI{20}{\kilo\electronvolt}) in the vacuum chamber setup. The average energy resolution of all pixels is~$\SI{246}{\electronvolt}$~FWHM (energy filter rise time of~\SI{2}{\micro\second}).}
\label{fig:pixelmap_fwhm}
\end{figure*}

For the following analyses of detector S0-166-4 in the monitor spectrometer, only the pixels with improved energy resolution values are taken into account, i.e.~pixels with the increased entrance window thickness are neglected. An average energy resolution of~\SI{336.7}{\electronvolt}~FWHM at~\SI{30.8}{\kilo \electronvolt} was obtained for the L-32 line of $^{\text{83m}}$Kr. The energy resolution of the individual pixels varies by at most~\SI{9}{\electronvolt} or approximately~\SI{3}{\percent} when compared to the average energy resolution. For completeness, the average energy resolution of the entire detector is~\SI{350.5}{\electronvolt}~FWHM at~\SI{30.8}{\kilo \electronvolt}. 
\\\\
For the detector \mbox{S0-166-6}, the cleaning procedure described above has been deployed during the detector assembly. The energy resolution for monoenergetic electrons with an energy of~\SI{20}{\kilo\electronvolt} is shown in figure~\ref{fig:pixelmap_fwhm_electron_tum}. The measurement was performed in the electron gun vacuum chamber setup, and an average energy resolution of~\SI{246}{\electronvolt}~FWHM was achieved. The energy resolution of the individual pixels varies by~\SI{15}{\electronvolt} or \SI{6}{\percent} compared to the average energy resolution. These results indicate a homogeneous performance of the detector response of the individual pixels. This is beneficial to reduce the complexity of the analysis for the keV sterile neutrino search with KATRIN by combining the measured energy spectrum of multiple pixels. Approximately, the combination of various pixels with slightly different responses can be described by an additional broadening of the spectrum. Sensitivity studies show that the energy resolution is not a limiting factor in the sterile neutrino search~\cite{Mertens_2015}. Currently, dedicated sensitivity studies are performed to estimate the impact of the varying detector response of different pixels and to determine the optimal way to combine pixels.
\noindent

\subsection{Energy dependence of the resolution}\label{ch:energy_resolution}
\begin{figure}[!t]
\centering
\includegraphics[width=1.0\linewidth]{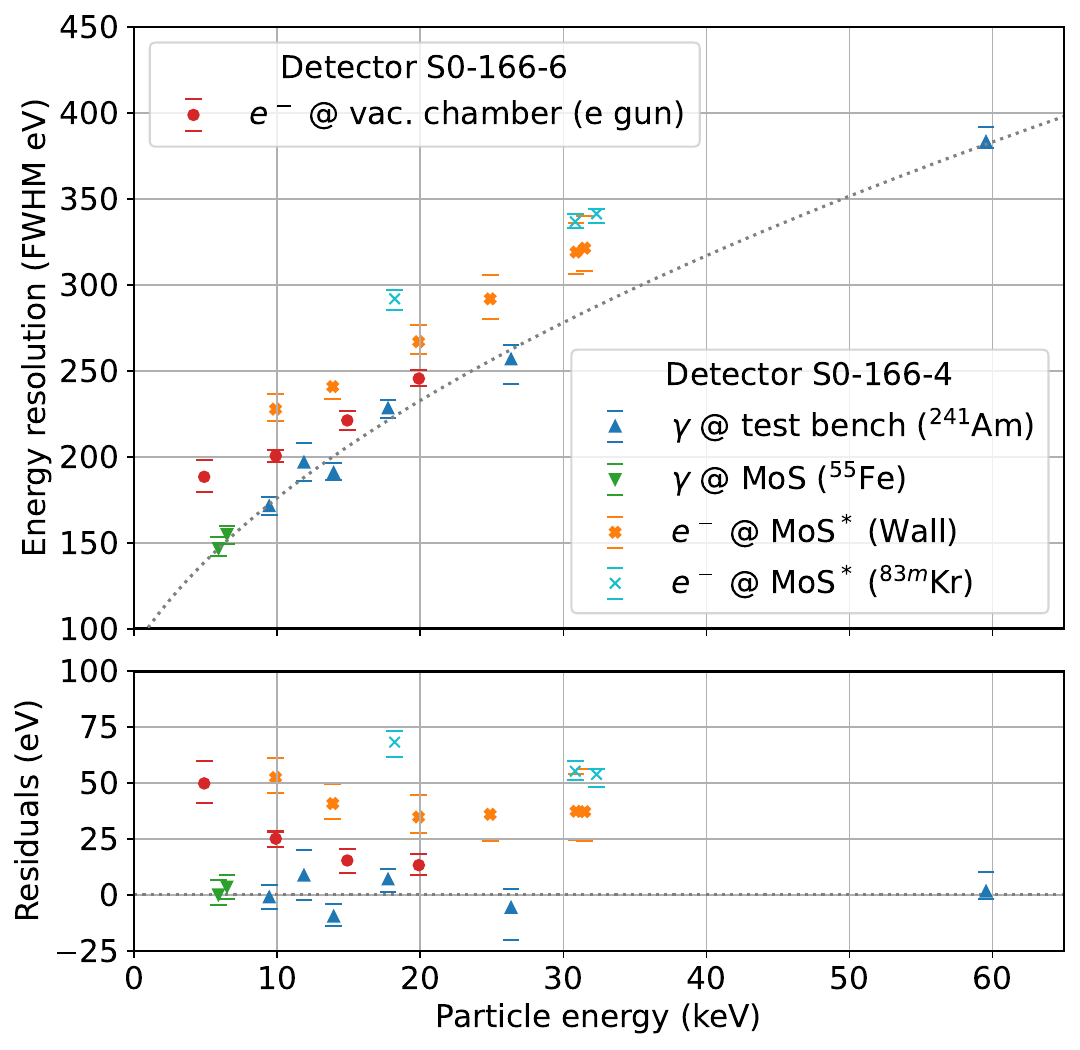}
\caption{Measured energy resolution as a function of the incoming particle energy for the different x-ray and electron sources. The horizontal bars of each data point indicate the minimal and maximal values of all pixels. The grey dotted line corresponds to the scenario of pure Fano statistics with an electronic noise of~$n_{\mathrm{ENC}}=9.9~e_{\text{rms}}$. For the electron data in the monitor spectrometer, only pixels without additional entrance window effects were taken into account~(indicated by the $^{\star}$). The bottom plot shows the residuals with respect to pure Fano statistics.}
\label{fig:fwhm_vs_energy}
\end{figure}
%
The energy resolution obtained in the different setups and with different sources as a function of energy is shown in figure~\ref{fig:fwhm_vs_energy}. For the x-rays and gamma rays, the data points are in good agreement with the expectation from Fano statistics when the measured noise of~$n_{\mathrm{ENC}}=9.9~e_{\text{rms}}$ of the detector system is taken into account. The lower and upper horizontal bars of each data point indicate the minimal and maximal measured energy resolution values of the individual pixels. An average energy resolution of~$\SI{228}{\electronvolt}$~FWHM at an x-ray energy of~\SI{17.8}{\kilo\electronvolt} was obtained. The energy resolution of the individual pixels varies by at most~\SI{11}{\electronvolt} or \SI{5}{\percent} when compared to the average value. 
\\\\
As shown in figure~\ref{fig:fwhm_vs_energy}, the energy resolution for electrons at a given energy is slightly worse than the one of x-rays. This can be related to entrance window effects leading to a low-energy tail as discussed above. The smaller the energy of the incident electrons, the higher is the effect on the energy resolution. This behavior is particularly visible for electrons with energies of \SI{5}{\kilo\electronvolt}. For energies above \SI{10}{\kilo\electronvolt}, the energy resolution is comparable to the energy resolution of x-rays. In general, for electrons with an energy of~\SI{20}{\kilo\electronvolt}, an average energy resolution of~$\SI{246}{\electronvolt}$~FWHM could be achieved. This result is compatible with the targeted energy resolution of \SI{300}{\electronvolt}~FWHM at \SI{20}{\kilo\electronvolt} for the keV sterile neutrino search with KATRIN.
\\\\
For the monitor spectrometer, the energy resolution analysis has been constrained to 29~pixels due to a contamination of the detector entrance window surface as discussed in~Section~\ref{ch:homogeneity}. The first measurements were performed with monoenergetic electrons from the spectrometer walls. The measured energy resolution is slightly worse compared to the one of a detector whose entrance window has been cleaned~(S0-166-6). In general, in the measurements with electrons from the spectrometer walls, an energy resolution of~$\SI{267}{\electronvolt}$ FWHM at~\SI{20}{\kilo\electronvolt} could be achieved. 
\\\\
For the electrons originating from the $^{\text{83m}}$Kr source, the measured energy resolution is slightly worse compared to the one of the wall electrons. This particularly holds at smaller energies. The difference can be attributed to additional scattering of the electrons inside the radioactive source resulting in an energy loss. An energy resolution of~\SI{292}{\electronvolt}~FWHM could be achieved for the electrons of the K-32 line at~\SI{18.2}{\kilo\electronvolt}. 

\subsection{Noise performance}\label{ch:noise}
\begin{figure}[!t]
\centering
\includegraphics[width=1.0\linewidth]{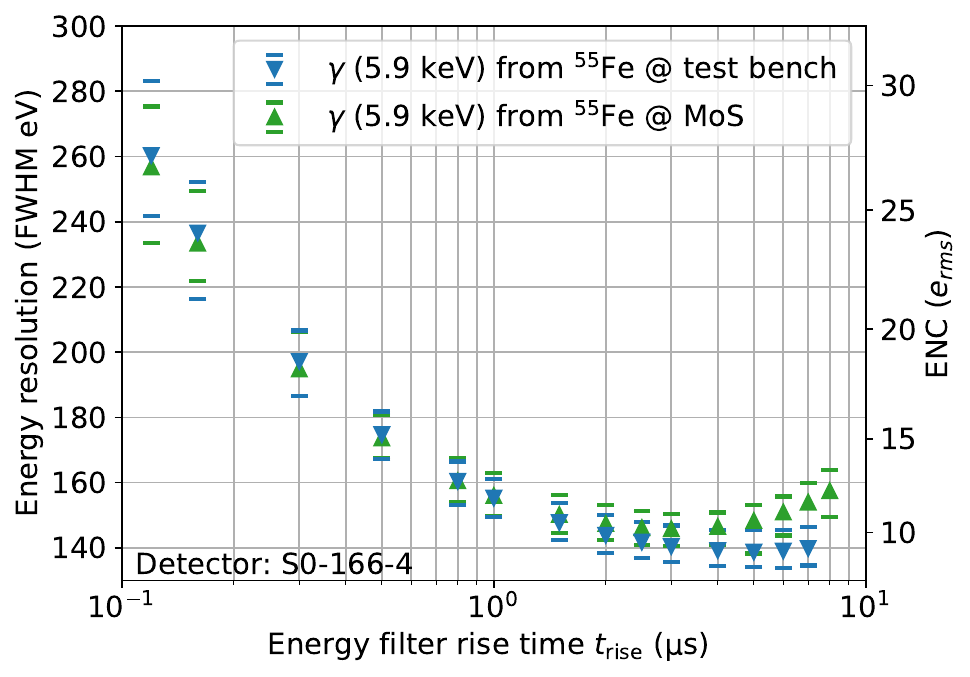}
\caption{Measured energy resolution of the $\text{Mn-K}_{\alpha}$ line ($^{55}$Fe source) as a function of the energy filter rise time. The plot shows the average values of all functional pixels for the measurements in the x-ray bench test setup~(blue points), and in the monitor spectrometer~(green points). The horizontal bars denote the minimal and maximal values of all pixels. For all data points, a filter flat top time of~$t_{\mathrm{top}}=\SI{0.3}{\micro\second}$ was used.}
\label{fig:noisecurve_comparison}
\end{figure}

\begin{figure}[!t]
\centering
\includegraphics[width=1.0\linewidth]{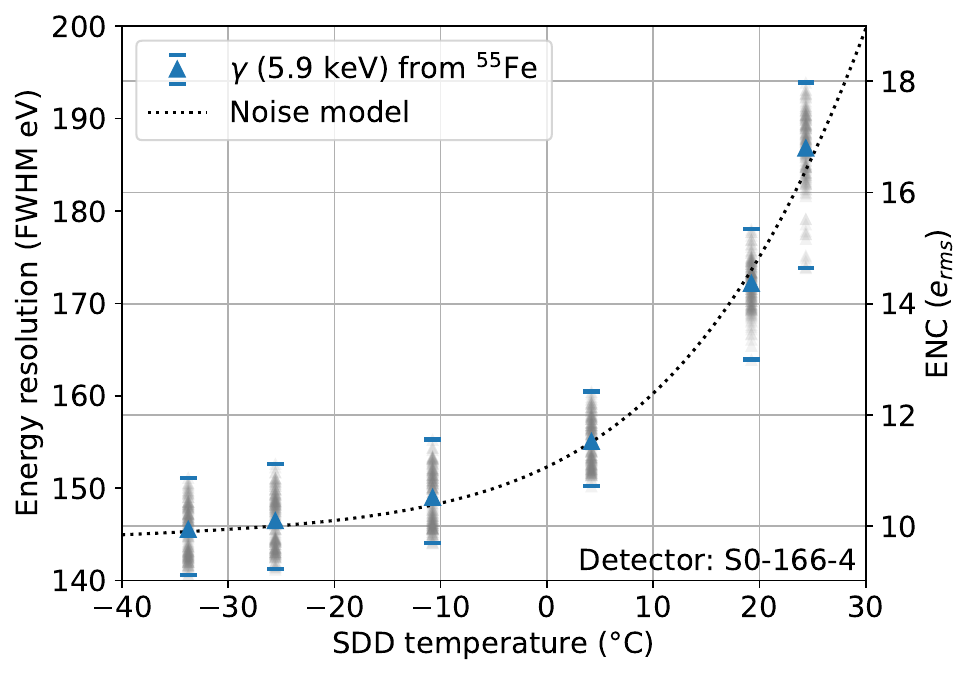}
\caption{Energy resolution of the $\text{Mn-K}_{\alpha}$ line (filter rise time of \SI{2}{\micro\second}) as a function of the detector temperature measured in the x-ray bench test setup. The average value of all functional pixels is indicated by the blue triangle. The horizontal bars denote the maximal and minimal values of all pixels. The values for the individual pixels at a given temperature are shown in grey.}
\label{fig:noise_vs_temperature}
\end{figure}

When searching for keV-scale sterile neutrinos, the TRISTAN detector system will measure at high rates of up to~\SI{100}{\kilo\countspersecond} per pixel. To limit the probability of pile-up events, short energy filter rise times of~$t_{\text{rise}}\leq\SI{2}{\micro\second}$ are required. However, at shorter filter rise times the energy resolution degrades, since the energy filter is more susceptible to the electronic noise. By determining the resolution at different filter rise times, the energy resolution can be optimized. To quantify it independent of the measured energy~$E$, the equivalent noise charge~$n_{\mathrm{ENC}}$~(ENC) of the detector system was calculated. It describes the number of electrons that would be created in the detector anode by a signal with the intensity of the root mean square~(rms) of the noise. To calculate~$n_{\mathrm{ENC}}$, the following equation was used:
\begin{equation}
\Delta E = 2\sqrt{2\ln2}\cdot\sqrt{n_{\mathrm{ENC}}^2\cdot w^2 + F\cdot w \cdot E}.
\end{equation}
Here, $w$ denotes the average energy necessary for the creation of an electron-hole pair~($w=\SI{3.63}{\electronvolt}$ for silicon), and $F=0.115$ the Fano factor which accounts for the non-Poissonian dispersion of the number of observed charge carriers~\cite{Fano_1980, Eggert_2004}.
\\\\
Figure~\ref{fig:noisecurve_comparison} shows measurements of the energy resolution of the~$\text{Mn-K}_{\alpha}$ line~($^{55}$Fe source) as a function of the energy filter rise time~$t_{\text{rise}}$ in the x-ray bench test setup and in the monitor spectrometer, respectively. In general, it can be observed that the resolution improves with increasing energy filter rise time. The best energy resolution with a value of~\SI{138.7}{\electronvolt}~FWHM was obtained for~$t_{\text{rise}}=\SI{5}{\micro\second}$ in the x-ray bench test setup. However, in this particular case, for count rates of up to~\SI{100}{\kilo cps}, more than~\SI{60}{\percent} of all events would have to be discarded due to pile-up. Decreasing the energy filter rise time to~$t_{\mathrm{rise}}=\SI{1}{\micro\second}$ leads to a slightly worsened energy resolution of~\SI{155.3}{\electronvolt}~FWHM~($n_\mathrm{ENC}=12.0~e_{\text{rms}}$) but retains more than~\SI{80}{\percent} of the events. As discussed in~\cite{Mertens_2015}, an energy resolution of up to~\SI{500}{\electronvolt} FWHM~($n_\mathrm{ENC}\approx50~e_{\text{rms}}$) does not limit the sensitivity of the keV~sterile neutrino search. Therefore, the energy filter rise time~$t_{\mathrm{rise}}$ could be reduced even further to minimize the pile-up probability. Currently, detailed studies using the results discussed above are carried out to optimize the filter setting, in order to maximize the sensitivity.
\noindent
\\\\
For filter rise times of~$t_{\mathrm{rise}}\leq\SI{1}{\micro \second}$, the average energy resolution values measured in the two different setups agree very well. This verifies the robustness of the detector against external electronic noise sources, particularly those that can be related to the complex measurement environment of the monitor spectrometer. For filter rise times of~$t_{\mathrm{rise}}\geq\SI{1}{\micro\second}$, the energy resolution measured in the monitor spectrometer is slightly worse than in the bench test setup. This difference might be attributed to an additional parallel noise component such as an increased leakage current. In particular, some light might have entered the monitor spectrometer vacuum vessel through a not perfectly sealed view port near the detector section. Nonetheless, for rise times~$t_{\mathrm{rise}}\geq\SI{1}{\micro\second}$, an average energy resolution better than~$\SI{160}{\electronvolt}$~FWHM was measured in both setups. This corresponds to an ENC of $n_{\mathrm{ENC}}<12.8~e_{\text{rms}}$.
\\\\
Compared to previous TRISTAN prototype detectors with 7 pixels (\SI{2}{\mm}~pixel diameter, external CUBE amplifiers)~\cite{Mertens_2019}, the energy resolution at short energy filter rise times is worse: For the prototype devices, an energy resolution below \SI{140}{\electronvolt}~FWHM at~$t_{\mathrm{rise}}=\SI{1}{\micro\second}$ was measured. The difference might be explained by the different readout electronics~(integrated JFET and ETTORE ASIC farther away vs stand-alone CUBE ASIC).
\\\\
One advantage of the SDD technology compared to other semiconductor detectors is its functionality at room temperature. This is due to the small leakage current on the order of~\SI{100}{\pico\ampere /\cm^2}. However, cooling down the detector to temperatures of about~\SI{-30}{\celsius} further reduces the leakage current to values below~\SI{2}{\pico\ampere /\cm^2}. To investigate the temperature dependence of the electronic noise, the TRISTAN detector was cooled down to different temperatures in the bench test setup. Measurements with an $^{55}$Fe source were performed and the energy resolution of the $\text{Mn-K}_{\alpha}$ line was determined as a function of the temperature, see figure~\ref{fig:noise_vs_temperature}. 
To approximate the temperature dependence, a simple exponential model was fitted to the mean energy resolution values~\cite{Eggert_2004}. Already for temperatures below~\SI{-10}{\celsius}, an average energy resolution smaller than~\SI{150}{\electronvolt}~FWHM ($n_{\mathrm{ENC}}=11~e_{\text{rms}}$) was measured. As indicated by the model, further decreasing the temperature of the detector down to values of~\SI{-34}{\celsius} slightly improves the energy resolution down to values of~\SI{144.5}{\electronvolt}~FWHM ($n_{\mathrm{ENC}}=9.9~e_{\text{rms}}$). Cooling down the detector module even further would not significantly improve the noise level. Therefore, this temperature was selected as the operating point for the measurements presented in this work.

\subsection{Linearity}\label{ch:linearity}
\begin{figure}[!t]
\centering
\includegraphics[width=1.0\linewidth]{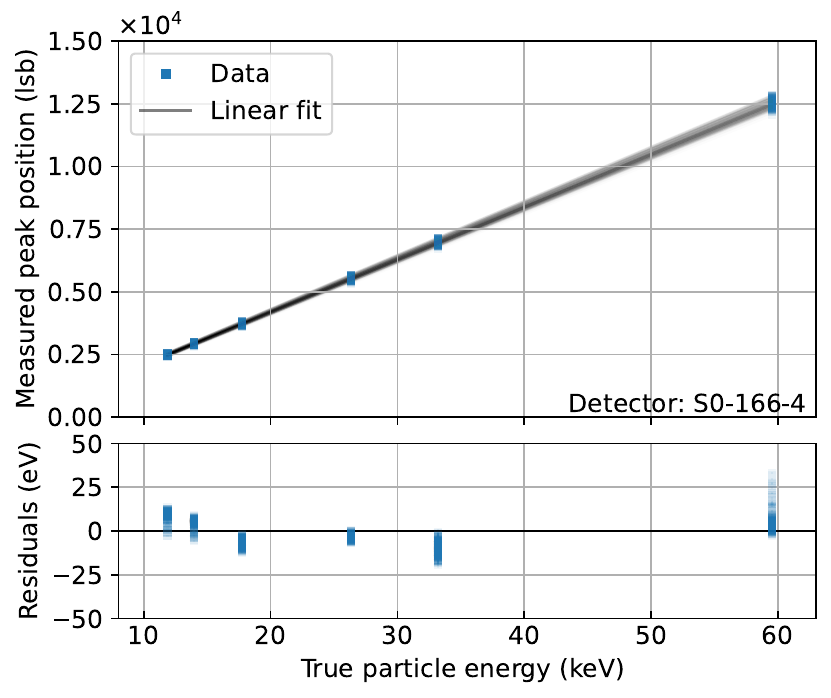}
\caption{Energy calibration of all pixels of the TRISTAN detector with an $^{241}$Am source in the x-ray bench test setup~(blue data points). The upper plot shows the relation between the measured peak position and the true particle energy. For each pixel, the relation is approximated with a linear function~(black lines). The lower plot shows the residuals.}
\label{fig:energy_calibration_am241}
\end{figure}

To investigate the linearity of the energy scale of the TRISTAN detector, a dedicated calibration measurement was performed using an~$^{241}$Am source in the x-ray bench test setup, see Section~\ref{ch:exp_setup_mpp}. Figure~\ref{fig:energy_calibration_am241} shows the measured peak positions~(in least significant bit units) as a function of the true energy values~(in keV units). A linear function is fitted to the data for each individual pixel to determine the calibration parameters~(slope and offset). In all cases, the measured peak position deviates less than~\SI{0.5}{\percent} from the true particle energy. It is worth mentioning that the calibration slope of the individual pixels varies by up to~\SI{4}{\percent}, requiring a separate energy calibration of each pixel. This is of major importance for the final TRISTAN detector system. Currently, detailed studies are ongoing, to assess the impact of different types of energy non-linearities on the final sensitivity to keV-scale sterile neutrinos. Previous studies have revealed that non-smooth non-linearities arising from ADC non-linearities can be a limiting systematic effect that needs to be mitigated~\cite{Dolde_2017}.

\subsection{Stability}\label{ch:stability}
\begin{figure}[!t]
\centering
\includegraphics[width=1.0\linewidth]{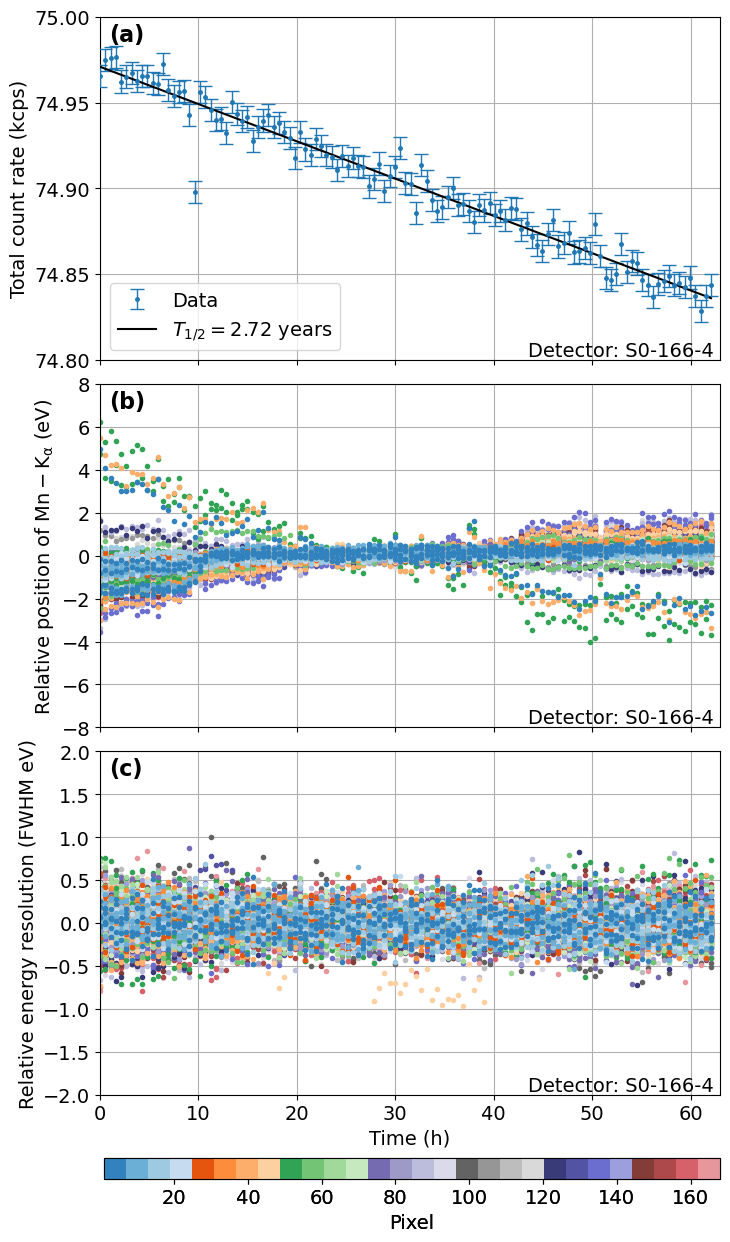}
\caption{Stability measurement with an $^{55}$Fe calibration source in the x-ray bench test setup. The individual pixels are indicated by the color given in the color bar. (a)~Count rate: Measured count rate summed over all pixels for events with an energy above~\SI{2}{\kilo\electronvolt} as a function of the measurement time. The decay of the count rate is fully compatible with the radioactive decay of $^{55}$Fe with a half-life of~$T_{1/2}=\SI{2.736}{years}$. (b)~Energy calibration: Relative shift of the peak position of the Mn-K$_{\alpha}$ line for every pixel and run. (c)~Energy resolution: Relative change of the energy resolution of the Mn-K$_{\alpha}$ line for every pixel and run.}
\label{fig:stability}
\end{figure}

To investigate the stability of the detector system, a measurement with a total duration of~63\,h (subdivided into several runs of~30\,min) was conducted in the x-ray test bench setup using an $^{55}$Fe source. The average temperature of the detector was~$\SI{-34.1}{\celsius}$ and varied by at most~\SI{1}{\kelvin} in the course of the measurement. 
\\\\
The measured total count rate in the detector for events with energies above \SI{2}{\kilo\electronvolt} is shown in figure~\ref{fig:stability}a. The decreasing rate is inherent to the decay of the $^{55}$Fe isotope. The relation was modeled with the exponential decay equation with one normalization parameter and the half-life~$T_{1/2}$. The best fit to the data was obtained for a half-life of~$T_{1/2}^{\star}=2.72^{+0.04}_{-0.05}\,\mathrm{years}$ which is in good agreement with the expected half-life of~$T_{1/2}=\SI{2.736}{~years}$~\cite{Pomme_2019}. To obtain an estimation of the rate stability of the detector, the residuals between the best fit values and the data were calculated. Normalizing the residuals by the total count rate per measurement point results in a maximal fluctuation of the count rate of~\SI{0.08}{\percent}.      
\\\\
In the next step, the stability of the energy calibration has been investigated. To this end, the relative shift of the peak position of the Mn-K$\alpha$ line 
\begin{equation}
\delta \mu_\text{px,run}=\mu_{\text{px, run}}-\overline{\mu}_{\text{px}}
\end{equation}
has been computed. Here, $\mu_{\text{px, run}}$ denotes the peak position extracted from a Gaussian fit for every pixel and run. Moreover, $\overline{\mu}_{\text{px}}$ is the average peak position of all runs for a given pixel. The relative shift of the peak position as a function of time is shown in figure~\ref{fig:stability}b. It can be observed that in the course of the measurement, for almost all pixels the peak position varies by less than~\SI{5}{\electronvolt}. Four pixels show a slightly higher drift of up to~\SI{10}{\electronvolt}. The origin of this drift is currently not clear and requires further investigations. Compared to the energy of the Mn-K$\alpha$ line at~\SI{5.9}{\kilo \electronvolt}, the maximal drift corresponds to less than \SI{0.2}{\percent} of the absolute energy.
\\\\
To investigate the stability of the energy resolution, the $\text{Mn-K}_{\alpha}$~line was fitted with a Gaussian function and the energy resolution~$\Delta E_{\text{px, run}}$ was extracted for all pixels and runs. Thereupon, the relative change of the energy resolution for each pixel 
\begin{equation}
\delta (\Delta E)_\text{px,run}=\Delta E_{\text{px, run}}-\overline{\Delta E}_{\text{px}}.
\end{equation}
has been calculated. Here, $\overline{\Delta E}_{\text{px}}$ describes the average energy resolution of all runs for each individual pixel. The stability of the energy resolution as a function of time is shown in figure~\ref{fig:stability}c. For all pixels, the energy resolution varies by at most~\SI{1}{\electronvolt} over the entire measurement period. Compared to the spread of the energy resolution between the individual pixels with~\SI{10}{\electronvolt}, this is very small.
\\\\
If all runs of a given pixel are combined prior to the energy calibration, the energy resolution for the individual pixels worsens by less than~\SI{1}{\electronvolt} compared to their best energy resolution in the individual runs. At least from the this point of view, it is therefore acceptable to perform measurements over multiple days without recalibration since the energy resolution is not a limiting factor for the keV~sterile neutrino search~\cite{Mertens_2015}. Currently, sensitivity studies are performed to assess the impact of the variation of the detector response to determine the optimal time window between energy calibration measurements.

\section{Conclusions and outlook}\label{ch:conclusions}
In this work we present the assembly, commissioning, and performance of the first TRISTAN detector modules. With different test stands and calibration sources the key parameters, such as energy resolution, pixel homogeneity, linearity, and time stability were investigated. This work shows that an average energy resolution of~\SI{246}{\electronvolt}~FWHM at~\SI{20}{\kilo\electronvolt} can be reached. This meets the requirement for the keV-scale sterile neutrino search at the ppm-level~\cite{Mertens_2015}. Moreover the pixel homogeneity and time stability over multiple days were demonstrated with a maximal spread of~\SI{15}{\electronvolt} between pixels (time stability of energy resolution~\SI{1}{\electronvolt}), which only has a small impact on the energy resolution, when combining pixels and runs. Furthermore, we measure an energy linearity at the sub-percent level. A follow-up study to \cite{Dolde_2017}, investigating the impact of realistic energy non-linearities is currently being performed. The detector was characterized in a semi-realistic MAC-E filter environment, i.e.~in vacuum and in a magnetic field. Here, the noise performance of the system was investigated in detail and the trade-off between pile-up events and energy resolution was characterized. Overall, an electronic noise of~$n_{\mathrm{ENC}}=9.9~e_{\text{rms}}$ was achieved~(energy filter rise time of~\SI{2}{\micro\second}) which is in good agreement with the results obtained under optimal lab conditions. This level is low enough such that the electronic noise does not affect the sensitivity of the search for keV-scale sterile neutrinos. Generally, the measured detector properties, such as energy resolution, pixel homogeneity, stability, and linearities, currently serve as a basis for a global sensitivity study, based on a full model of the measured tritium spectrum. As these sensitivity studies are still ongoing at the moment, we refer the reader to a future publication for the final assessment of detector-related systematic uncertainties and their impact on the final sensitivity.
\\\\ 
The next step in the development of the detector system is the assembly and characterization of an integrated system, consisting of multiple detector modules. Additionally, a high-luminosity electron gun is currently under developed and will be used to validate the results also at the envisioned rate of~\SI{100}{\kilo\countspersecond} per pixel. Finally, the preparations for the integration of the detector system into the KATRIN beam line, after its neutrino mass measurement campaign is completed end of 2025, are currently ongoing. 


\begin{acknowledgements}
We acknowledge the support of Helmholtz Association (HGF); Ministry for Education and Research BMBF (05A17PM3, 05A17PX3, 05A17VK2, 05A17PDA, 05A17WO3, 05A20VK3, 05A20PMA and 05A20PX3); Helmholtz Alliance for Astroparticle Physics (HAP); the doctoral school KSETA at KIT; Helmholtz Young Investigator Group (VH-NG-1055); Max Planck Research Group (MaxPlanck@TUM); Deutsche Forschungsgemeinschaft DFG (Research Training Group grant nos. GRK 1694 and GRK 2149); Graduate School grant no. GSC 1085-KSETA,  SFB-1258, and Excellence Cluster ORIGINS in Germany; Ministry of Education, Youth and Sport (CANAM-LM2015056, LTT19005) in the Czech Republic; the Department of Energy through grants DE-FG02-97ER41020, DE-FG02-94ER40818, DE-SC0004036, DE-FG02-97ER41033, DE-FG02-97ER41041, DE-SC0011091 and DE-SC0019304; and the Federal Prime Agreement DE-AC02-05CH11231 in the USA. This project has received funding from the European Research Council (ERC) under the European Union Horizon 2020 research and innovation programme (grant agreement no. 852845). We thank the computing cluster support at the Institute for Astroparticle Physics at Karlsruhe Institute of Technology, Max Planck Computing and Data Facility (MPCDF), and National Energy Research Scientific Computing Center (NERSC) at Lawrence Berkeley National Laboratory.
\end{acknowledgements}
\printbibliography
\end{document}
\typeout{get arXiv to do 4 passes: Label(s) may have changed. Rerun}